\documentclass[lettersize,journal]{IEEEtran}
\ifCLASSINFOpdf
\else
\fi

\usepackage{tabularx}
\usepackage{graphics}
\usepackage{bbm}
\newtheorem{theorem}{{Theorem}}

\usepackage{stfloats}

\usepackage{CJK}
\usepackage{algorithm}
\usepackage{algorithmic}
\usepackage{amsmath}
\usepackage{amssymb}
\usepackage{psfrag}
\usepackage{graphicx}
\usepackage{epstopdf}
\usepackage{multirow}
\usepackage{stfloats}
\usepackage{cite}
\usepackage{subcaption}
\usepackage{color}
\usepackage[normalem]{ulem}
\usepackage{arydshln}
\usepackage{enumerate}
\usepackage{bbm}
\usepackage{verbatim}
\usepackage{gensymb}

\newcommand{\bm}[1]{\mbox{\boldmath{$#1$}}}
\title{On the Capacity Region of Reconfigurable Intelligent Surface Assisted Symbiotic Radios}

\author{Qianqian Zhang, Hu Zhou, Ying-Chang Liang, \emph{Fellow, IEEE},
Sumei Sun, \emph{Fellow, IEEE}, \\
Wei Zhang, \emph{Fellow, IEEE},  and
 H. Vincent Poor, \emph{Fellow, IEEE} \\

\thanks{
Part of this work has been presented in IEEE ICC 2023~\cite{zhang2023ICC}.
Q.~Zhang, H.~Zhou, and Y.-C. Liang are with University of Electronic Science and Technology of China (UESTC), Chengdu 611731, China (e-mails: qqzhang\_kite@163.com, huzhou@std.uestc.edu.cn, and liangyc@ieee.org).
S. Sun is with Institute for Infocomm Research, A*STAR, Singapore (e-mail: sunsm@i2r.a-star.edu.sg).
W. Zhang is with University of New South Wales, Sydney, NSW 2052, Australia (e-mail: wzhang@ee.unsw.edu.au).
H. V. Poor is with the Department of Electrical and Computer Engineering, Princeton University, Princeton, NJ 08544 USA (e-mail: poor@princeton.edu).
}}

\begin{document}

\maketitle

\vspace{-3em}
\begin{abstract}

In this paper, we are interested in a reconfigurable intelligent surface (RIS)-assisted symbiotic radio (SR) system, where an RIS assists a primary transmission by passive beamforming and simultaneously acts as an information transmitter by periodically adjusting its reflection coefficients. The above RIS functions innately enable a new multiplicative multiple access channel (M-MAC), where the primary and secondary signals are superposed in a multiplicative manner. To pursue the fundamental performance limits of M-MAC, we focus on the characterization of the capacity region for such a system when the direct link is blocked. Due to the reflection nature of RIS, the signal transmitted
from the RIS should satisfy a passive reflection constraint. We consider two types of passive reflection constraints, one for the case that
only the phases of the RIS can be adjusted, while the other for the case that both the amplitude and the phase can be adjusted. 
Under the passive reflection constraints at the RIS as well as the average power constraint at the primary transmitter (PTx), 
we characterize the capacity region of RIS-assisted SR. It is observed that: 
1) the number of sum-rate-optimal points on the boundary of the capacity region is infinite;
2) for the rate pairs with the maximum sum rate, the optimal distribution of the amplitude of the primary signal is a continuous Rayleigh distribution, 
while for the remaining rate pairs on the capacity region boundary, the optimal amplitude of the primary signal is discrete;
3) the adjustment of the amplitude for the RIS can enlarge the capacity region as compared to the phase-adjusted-only case.


\end{abstract}


 \begin{IEEEkeywords}

Symbiotic radio, reconfigurable intelligent surface, capacity region, multiplicative multiple access channel.
 \end{IEEEkeywords}

\section{Introduction}

\IEEEPARstart{W}{ireless} communication systems continue to strive for ever higher data rates and more access devices. It is envisioned that the sixth generation (6G) mobile networks will accommodate up to $1$ terabyte per second (Tb/s) peak data rates and more than $10^7$ connections per square kilometer~\cite{matti2019key,giordani2020toward}. This is a particularly challenging task for wireless communication due to the availability of limited power and spectrum resources~\cite{xiaohutowards,ge2023deep}.
Symbiotic radio (SR) has emerged as a promising technology to overcome the challenges of spectrum scarcity and high power consumption due to its spectrum- and power-sharing natures. Specifically, in SR, a passive secondary transmitter (STx) modulates its information over the signal emitted from a primary transmitter (PTx) by periodically adjusting its reflecting coefficient~\cite{liang2020symbiotic,liu2013ambient,Boyer2014Backscatter}.
This modulation scheme innately enables the STx to transmit information without dedicated spectrum and energy resources, 
thereby yielding high spectral and energy efficiencies~\cite{yang2018modulation}. In return, the secondary transmission provides multipath gain to the primary transmission, thereby yielding mutual benefits between the two transmissions~\cite{zhang2022mutualistic}. Due to such mutualistic features and high spectral and energy efficiencies, SR has attracted extensive attention from both academia and industry recently~\cite{chen2020vision,9178307}.

In SR, a number of studies have been carried out on the mutualistic relationship, receiver design, and resource allocation schemes. Particularly, a mutualistic condition of SR through which both primary and secondary transmissions can benefit each other is analyzed in~\cite{zhang2022mutualistic} from the perspective of bit error rate (BER) performance. Various types of detectors and their corresponding BERs are studied in~\cite{yang2018cooperative}. In~\cite{chen2020stochastic}, stochastic optimization techniques are used to design the transceiver for the SR system. In~\cite{zhang2019constellation} and~\cite{guo2019exploiting}, a constellation learning detector and an iterative detector are designed, respectively, to recover signals transmitted from the STx when the secondary receiver does not have the pilot information of the primary transmission. As for resource allocation schemes, beamforming vectors of the PTx are designed in~\cite{long2019symbiotic} and~\cite{chu2020resource} under infinite-block length and finite-block length scenarios, respectively. A transmit power minimization problem is studied in~\cite{10499212} for a multi-user
SR system to design the beamforming vectors of multiple PTxs.


To address the double fading effect associated with SR, reconfigurable intelligent surface (RIS) has been proposed to act as the STx for SR.
Specifically, in an RIS-assisted SR system, the RIS delivers information by periodically adjusting its reflection coefficients, which is an extension of backscatter modulation~\cite{zhang2021reconfigurable}. Besides, RIS can also deliver information by spatial modulation~\cite{ma2020large,liang2022backscatter} and symbol level precoding~\cite{liu2019symbol,liang2022backscatter}. 
Meanwhile, a novel modulation scheme is proposed for the RIS in~\cite{zhou2023modulation}, where the phase-shift matrix of the RIS
is divided into two components, one to assist primary transmission and the other to deliver messages.
To balance the performance between primary and secondary transmissions, an RIS partitioning scheme is developed in~\cite{10336749}, 
in which the RIS is partitioned into two sub-surfaces to assist primary transmission and to transmit signals, respectively.
The effect of the number of reflecting elements of the RIS on mutualistic mechanisms is analyzed in \cite{wang2023mutualistic}.
Moreover, in~\cite{karasik2021single}, the capacity of RIS-assisted SR is derived under discrete transmitted constellation points. 
The degree-of-freedom of RIS-assisted SR is characterized in~\cite{cheng2021degree} when PTx and RIS transmit multiple data streams.
Furthermore, there are some studies on the joint design of the transmit beamforming at the PTx and the reflection coefficients at the RIS.
Particularly, for RIS-assisted MIMO SR, the transmit power minimization problem is investigated in~\cite{zhang2021reconfigurable}. 
For RIS-assisted MISO SR, BER minimization, transmit power minimization, and secondary transmission rate maximization problems 
are studied in~\cite{hua2021novel},~\cite{zhou2022cooperative}, and \cite{zhang2020sr}, respectively.

Despite the above notable advancements, it is still challenging to address the fundamental performance limits of RIS-assisted SR 
systems due to the following two reasons. On one hand, the primary and secondary signals in SR are superposed in a multiplicative manner, which forms a new multiplicative multiple access channel (M-MAC)~\cite{liu2018backscatter}.
Although there are extensive studies on additive multiple access channel (MAC)~\cite{tse2005fundamentals} over the last few decades,
the results of additive MAC can not be applied to M-MAC directly.
On the other hand, the RIS is a passive device whose input should satisfy a passive reflection constraint. That means the peak amplitude of the signal transmitted from the RIS is limited, and thus the widely studied channel capacities under average power constraints are no longer applicable. 

As for M-MAC, in~\cite{cover1999elements}, the capacity region of a binary M-MAC without noise has been studied. The capacity region of the M-MAC with noise is characterized under average power constraints in~\cite{pillai2011capacity}. It is worth noting that the capacity region in the above two cases is a triangle.
Meanwhile, there are several studies on single-user channel capacity with peak power constraints. In this scenario, the most interesting results are due to Smith~\cite{smith1971information}, which indicate that the capacity-achieving distribution is discrete with a finite number of mass points. In~\cite{shamai1995capacity}, a peak power limited quadrature Gaussian channel is considered, in which the optimal input distribution is supported on a finite number of concentric circles, i.e., discrete amplitude and uniform independent phase. The $n$-dimensional vector Gaussian noise channel is investigated under peak power constraints in~\cite{dytso2019capacity}, where the optimal input distribution is geometrically characterized by concentric spheres.
Moreover, there are some studies on the characterization of capacity region under peak power constraints over an additive MAC~\cite{mamandipoor2014capacity,ozel2012capacity}, which show that any point on this capacity region boundary can be achieved by discrete input distributions of finite support.
Despite of all of this, when M-MAC meets passive reflection constraints, the optimal input distributions and the capacity region of 
RIS-assisted SR are still unknown.

In this paper, we are interested in the characterization of capacity region for an RIS-assisted SR system. 
Since the presence or absence of a direct link will lead to different capacity regions, we focus on the blocked direct link case in this paper.
We consider two types of passive reflection constraints, one for which only the phases of the RIS can be adjusted, 
while the other for which both the amplitude and the phase can be adjusted.
To pursue the fundamental performance limits of such a system, we first analyze optimal distributions of transmitted signals under average power constraint at the PTx and passive reflection constraint at the RIS, and then derive the maximum achievable rates for both primary and secondary transmissions as well as their maximum sum rate. Then, we derive the optimal distributions of transmitted signals for the rate pairs on the boundary of
the capacity region and characterize the capacity region of RIS-assisted SR with the two types of passive reflection constraints. 
The main theoretical results are summarized as follows: 1) the secondary transmission achieves the maximum rate when the PTx transmits a sinusoidal 
signal with the blocked direct link; 2) when the amplitude of the RIS can be adjusted, the capacity-achieving distribution of the secondary signal is geometrically characterized by
concentric circles; 3) the primary transmission achieves the maximum rate when the RIS only assists the primary transmission and the PTx 
transmits Gaussian signals; 4) the number of sum-rate-optimal points on the capacity region boundary is infinite which is achieved by the joint
design of distributions of transmitted signals; 5) for the rate pairs with the maximum sum rate, the optimal distribution of the amplitude of the primary signal is a continuous Rayleigh distribution,
while for the remaining rate pairs on the boundary, the optimal amplitude of the primary signal is discrete; 6) the adjustment of the amplitude for the RIS can increase the
degree-of-freedom of the secondary transmission and enlarge the capacity region.
In a nutshell, the main contributions of this paper are summarized as follows.
\begin{itemize}
  \item We derive the optimal distributions of transmitted signals to achieve the maximum achievable rates for both primary and secondary transmissions as well as their maximum sum rate.
  \item We derive the optimal distributions of transmitted signals for the rate pairs on the boundary of capacity region and characterize 
        capacity region of RIS-assisted SR.
  \item Extensive numerical results are provided to evaluate the performance of RIS-assisted SR and demonstrate the features of M-MAC.
\end{itemize}

The rest of the paper is organized as follows. In Section II, we establish the system model and represent the capacity region mathematically for RIS-assisted SR. The capacity regions for RIS-assisted SR for different passive reflection constraints are characterized in Sections III
and IV, respectively. The performance evaluation is presented in Section V. Finally, the paper is concluded in Section VI.

Notations used in this paper are listed as follows. The lowercase and boldface lowercase $x$ and $\mathbf x$ denote a scalar variable (or constant) and a vector, respectively. $\mathcal C \mathcal N(\bm {\mu}, \bm\Sigma)$ denotes the complex Gaussian distribution with mean $\bm\mu$ and variance $\bm \Sigma$. $U(a,b)$ denotes the uniform distribution between the interval $(a,b)$.
$\mathbf x^H$ denotes the conjugate transpose of $\mathbf x$. $\mathbb{E}(\cdot)$ denotes the statistical expectation. $\mathrm{cl}(\mathrm{conv}(\mathcal A))$ denotes the closure of the convex hull of a subset $\mathcal A$. $\mathrm{diag}(\mathbf x)$ denotes a diagonal matrix whose diagonal elements are given by the vector $\mathbf{x}$. $\delta(\cdot)$ stands for the Dirac-$\delta$ function. $\mathcal L(\cdot)$ and $\mathcal L^{-1}(\cdot)$ denote Laplace transform and inverse Laplace transform, respectively.

\section{RIS-assisted SR}

In this section, we will describe the system model of RIS-assisted SR and illustrate mathematically the definition of capacity region for the RIS-assisted SR system.

\subsection{System Model}

The RIS-assisted SR system consists of a single-antenna PTx, a single-antenna cooperative receiver (C-Rx), and an RIS equipped with $K$ reflecting elements. The PTx is an active device with maximum average power $P$, whose transmitted signal $X_1 \in \mathbb C$ satisfies an average power constraint $\mathbb{E}[|X_1|^2]\leq P$.
The RIS is a passive device, which transmits signal $X_2\in \mathbb C$ by periodically adjusting its reflection coefficients, i.e., $\bm \phi = [\phi_1,\cdots,\phi_K]^T$. The mapping between $\bm \phi$ and $X_2$ is given by $\bm \phi = \rho \bm \varphi X_2$, where
$\rho$ is reflection efficiency with $\rho \leq 1$ and $\bm \varphi = [\varphi_1,\cdots,\varphi_K]^T$ is used for passive beamforming. 

It is worth noting that the reflection coefficients of RIS should meet the passive reflection constraint due to the reflection nature of RISs.
Typically, the reflection coefficients follow a phase shift model, where only the phase of each reflection coefficient is adjustable. 
In this case, to normalize $X_2$, we have $|\varphi_k| = 1,\forall k$ and $|X_2| = 1$.
On the other hand, we notice that some works point out that the amplitude of each reflection coefficient can also be 
adjusted~\cite{liu2019intelligent,tang2020mimo,zhao2021exploiting}, yielding the passive reflection constraint $|X_2| \leq 1$.
As will hereafter be seen, the adjustment of the amplitude of reflection coefficients
will increase the degree-of-freedom of the secondary transmission. Thus, in this paper, we will characterize capacity regions with 
constraint $|X_1| = 1$ and constraint $|X_1| \leq 1$, respectively.

\begin{figure}[t]
\centering
\includegraphics[width=.7\columnwidth] {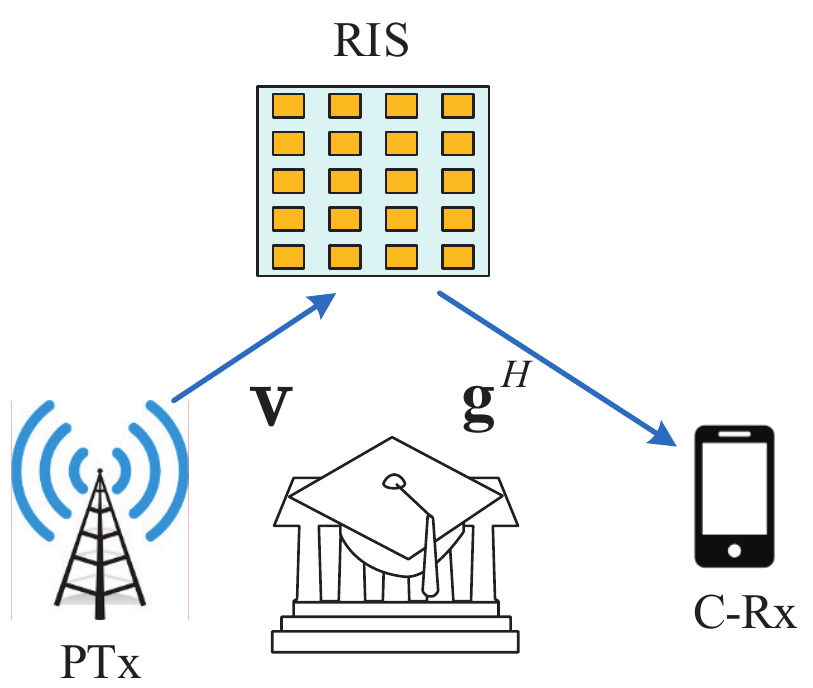}
\caption{System model.}
\label{system model}
\end{figure}

Since capacity regions with and without the direct link from PTx to C-Rx for RIS-assisted SR are different, in this paper, we focus on the case 
where the direct link is blocked. The capacity region characterization with the direct link for RIS-assisted SR will be investigated in 
another paper.
As shown in Fig.~\ref{system model}, 
denote by $\mathbf v = [v_1,\cdots,v_K]^T\in \mathbb C^{K\times 1}$ the channel coefficient from the PTx to the RIS, and by $\mathbf g^H = [g_1,\cdots,g_K]\in \mathbb C^{1\times K}$ the channel coefficient from the RIS to the C-Rx. The received signal $Y$ at the C-Rx is given by
\begin{align}
 \label{eq:system model}
{Y} &=   \rho\mathbf g^H\bm\Psi\mathbf v X_1X_2 + {Z}\nonumber\\
 &=  hX_1X_2 + {Z},
\end{align}
where $ h\triangleq\rho\mathbf g^H\bm\Psi\mathbf v$, $\bm\Psi = \mathrm{diag}( \bm \varphi)$, and $Z\sim\mathcal{C}\mathcal{N}( 0,\sigma^2 )$ is the additive white Gaussian noise at the C-Rx. We assume the channel state information (CSI) is perfectly known to the transmitters and the receiver. 

\subsection{Definition of Capacity Region}\label{sec:region}

Denote by $R_1$ and $R_2$ the achievable rates of primary and secondary transmissions, respectively.
Under a low enough average error probability, the capacity region of M-MAC is the closure of the convex hull of all rate pairs $(R_1,R_2)$, which is denoted by $\mathcal C_{\textrm{M-MAC}}$~\cite{cover1999elements}.
Following standard approaches in~\cite{cover1999elements}, the capacity region of M-MAC in RIS-assisted SR is characterized as
\begin{align}
 \label{eq:system model}
&\mathcal C_{\textrm{M-MAC}}(R_1,R_2)\nonumber \\
= &\mathrm{cl}\left(\mathrm{conv}\left(\bigcup_{\bm\varphi, f_1(X_1), f_2(X_2)}\!\!\!\mathcal R(R_1,R_2;\bm \varphi,X_1, X_2)\!\right)\right),
\end{align}
where $f_1(X_1)$ and $f_2(X_2)$ are the probability density functions (PDFs) of $X_1$ and $X_2$, respectively, and
$\mathcal R(R_1, R_2;\bm \varphi,X_1, X_2)$ is the set of all tuples $(R_1,R_2)$ satisfying:
\begin{align}
R_1 & \leq I(X_1;Y|X_2,Q), \label{eq:system model formula1}\\
R_2 & \leq I(X_2;Y|X_1,Q),\label{eq:system model formula2}\\
R_1+R_2 & \leq I(X_1,X_2;Y,Q) \label{eq:system model formula3}.
\end{align}
Note that $Q$ is the time-sharing random variable, which will be discussed in the next section.

\vspace{-0.5em}

\section{Capacity Region Characterization with Constraint $|X_2|=1$}\label{sec:Rate}

In this section, we will characterize the capacity region of RIS-assisted SR with constraint $|X_2| = 1$. First, based on the definition of capacity region, we characterize the maximum achievable rates of primary and secondary transmissions as well as their maximum sum rate. Then, we describe rate pairs on the boundary of capacity region.

\subsection{Maximum Achievable Rate of Primary Transmission}\label{sec:R1}

The maximum achievable rate of the primary transmission is denoted by $C_1$, which can be obtained by solving the following problem:
\begin{align}
\mathop {\max }\limits_{\bm \varphi, f_1(X_1), f_2(X_2)}&I(X_1;Y|X_2)\\
\textrm{s.t.}~\qquad& \mathbb{E}[|X_1|^2]\leq P \label{eq:c1}\\
&|X_2| = 1\label{eq:c2}\\
&|\varphi_k| = 1,\forall k. \label{eq:c3}
\end{align}
Given $\bm \varphi$, for any distributions $f_1(X_1)$ and $f_2(X_2)$, we have
\begin{align}
I(X_1;Y|X_2) &=\int_{X_2}f_2(X_2)I(X_1;Y|X_2 = x_2)dx_2\nonumber\\
&{\leq} ~\mathop {\max }\limits_{x_2} I(X_1;Y|X_2 = x_2),
\end{align}
since the average is less than the maximum. That means when the RIS purely assists the primary transmission, the
primary transmission rate achieves the maximum.

Under average power constraints, Gaussian distribution can achieve the maximal entropy over all distributions~\cite{cover1999elements}. Thus, when $X_1 \sim \mathcal{CN}(0,P)$, the primary transmission can achieve the maximum rate, which is given by
\begin{equation}
 \label{eq:pri2}
C_1 = \mathop {\max }\limits_{\bm \varphi, x_2}~\log\left(1+\frac{P|hx_2|^2}{\sigma^2}\right).
\end{equation}

From \eqref{eq:pri2}, when $|hx_2|^2$ is maximized, the primary transmission can achieve the maximum rate, whose solution is $\varphi_k x_2 = e^{\vartheta-j\mathrm{arg}(v_kg_k)}$ and $\vartheta$ is an arbitrary value from $-\pi$ to $\pi$. By defining $\tilde{h}  =\sum_{k=1}^{K}\rho|v_kg_k|$, we have
\begin{equation}
 \label{eq:primary}
C_1 = \log\left(1+\frac{P\tilde{h}^2}{\sigma^2}\right).
\end{equation}

\subsection{Maximum Achievable Rate of Secondary Transmission}\label{sec:R2}

In this section, we will provide the maximum achievable rate of the secondary transmission, which is denoted by $C_2$ and can be obtained 
by solving the following problem:
\begin{align*}
\mathop {\max }\limits_{\bm \varphi, f_1(X_1), f_2(X_2)}&I(X_2;Y|X_1)\\
\textrm{s.t.}~\qquad& \eqref{eq:c1}, \eqref{eq:c2}, \textrm{and }\eqref{eq:c3}.
\end{align*}

Based on the definition of mutual information, we have
\begin{align}\label{eq:ne11}
I(X_2;Y|X_1) = \int_{X_1}f_1(X_1)I(X_2;Y|X_1 = x_1)d x_1.
\end{align}
According to~\cite{wyner1966bounds}, the optimal distribution of $f_2(X_2)$ to maximize $I(X_2;Y|X_1 = x_1)$ is characterized by a uniformly distributed independent phase from $-\pi$ to $\pi$, i.e., $X_2 = e^{j\theta_2}$ and $\theta_2 \sim U[-\pi,\pi)$.
\begin{theorem}\label{theorem:1}
The mutual information $I(X_2;Y|X_1 = x_1)$ can be calculated by
\begin{equation}
 \label{eq:sec1}
I(X_2;Y|X_1 = x_1) = -\int_{0}^{+\infty}\frac{2{r}\kappa({r})}{\sigma^2} \log\left(\kappa({r})\right) d{r}-\log(e),
\end{equation}
where $\kappa(r) =  \exp\left(-\frac{{r}^2+|hx_1|^2}{\sigma^2}\right)I_0\left(\frac{2{r}|hx_1|}{\sigma^2}\right)$ and $I_0(x) = \frac{1}{2\pi}\int_{-\pi}^{\pi}e^{x\cos\theta}d\theta$.
\end{theorem}
\begin{IEEEproof}
Please see Appendix \ref{proof:capacity}.
\end{IEEEproof}

An asymptotic behavior of \eqref{eq:sec1} is given by~\cite{wyner1966bounds}
\begin{align}\label{eq:asy}
\!\!I(X_2;Y|X_1 \!=\! x_1)\! \approx \! \left\{
\begin{array}{cl}
  \! \frac{1}{2}\log\left(\frac{4\pi|hx_1|^2}{e\sigma^2}\right),&\!\!{\frac{|hx_1|^2}{\sigma^2}\gg1},    \\
  \!\frac{|hx_1|^2}{\sigma^2},&\!\!{\frac{|hx_1|^2}{\sigma^2}\ll1}.
  \end{array}
  \right.
\end{align}

From~\eqref{eq:asy}, one can find that $I(X_2;Y|X_1 = x_1)$ is a monotonically increasing function of the receive signal-to-noise ratio (SNR), i.e., $\frac{|hx_1|^2}{\sigma^2}$. Thus, $\bm \varphi$ can be designed to maximize $|h|^2$, whose solution is $\varphi_k = e^{j(\vartheta-\mathrm{arg}(v_kg_k))}$, where $\vartheta$ is an arbitrary value from $-\pi$ to $\pi$.

We next discuss the optimal distribution of $X_1$ to maximize $I(X_2;Y|X_1)$.
From~\eqref{eq:asy}, it is obvious that $I(X_2;Y|X_1 = x_1)$ a concave function over $|x_1|^2$. Since the phase of $X_1$ does not affect $I(X_2;Y|X_1 = x_1)$ from~\eqref{eq:sec1} and~\eqref{eq:asy}, then based on Jensen's inequality, we have
\begin{align}
I(X_2;Y|X_1) =& ~\mathbb E_{x_1}I(X_2;Y|X_1 = x_1)\nonumber\\
{\leq}& ~I(X_2;Y||X_1|^2 = P).
\end{align}
Thus, when $|X_1|^2 = P$, the maximum rate is achievable for the secondary transmission.
Note that the phase of $X_2$ follows a uniform distribution over $[-\pi,\pi)$. To avoid information ambiguity, the PTx is required to
transmit a sinusoidal signal without phase information to maximize the achievable rate of the secondary transmission.

Given $f_2(X_2)$, $X_1 = \sqrt{P}$, and $\varphi_k = e^{j(\vartheta-\mathrm{arg}(v_kg_k))}$, the maximum achievable rate of the secondary transmission is given by
\begin{equation}
 \label{eq:secondary1}
C_2 = -\int_{0}^{+\infty}\frac{2{r}\underline{\kappa}({r})}{\sigma^2} \log\left(\underline{\kappa}({r})\right) d{r}-\log(e),
\end{equation}
where $\underline{\kappa}(r) = \exp\left(-\frac{{r}^2+{P}\tilde h^2}{\sigma^2}\right)$ $I_0\left(\frac{2{r}\sqrt{P}\tilde h}{\sigma^2}\right)$.
An asymptotic behavior of \eqref{eq:secondary1} is given by~\cite{wyner1966bounds}
\begin{align}\label{eq:asy1}
C_2  \approx \left\{
\begin{array}{cl}
   \frac{1}{2}\log\left(\frac{4\pi P\tilde{h}^2}{e\sigma^2}\right),\;\;&{\frac{P\tilde{h}^2}{\sigma^2}\gg1},    \\
  \frac{P\tilde{h}^2}{\sigma^2},\;\;&{\frac{P\tilde{h}^2}{\sigma^2}\ll1}.
  \end{array}
  \right.
\end{align}
From~\eqref{eq:secondary1} and $\eqref{eq:asy1}$, one can see that the PTx serves as an energy source for the RIS to transmit information, which is equivalent to a system where a transmitter transmits messages $\sqrt{P}e^{j\theta_2}$ with channel $\tilde h$, where $\theta_2 \sim
U[-\pi,\pi)$. Since under average power constraints, Gaussian distribution achieves the maximal entropy over all distribution, we know that the maximum achievable rate of the secondary transmission $C_2$ is less than $C_1$.

\subsection{Maximum Achievable Sum Rate}\label{sec:Rsum}

The maximum achievable sum rate of the primary and secondary transmissions is denoted by $C_{\textrm{sum}}$, which is obtained by
solving the following problem:
\begin{align*}
\mathop {\max }\limits_{\bm \varphi, f_1(X_1), f_2(X_2)} &I(X_1, X_2;Y)\\
\textrm{s.t.}~\qquad& \eqref{eq:c1}, \eqref{eq:c2}, \textrm{and }\eqref{eq:c3}.
\end{align*}

To characterize the maximum sum rate, we consider another transmission case where $(X, 1)$ are symbols transmitted at the PTx 
and RIS, respectively, with $ X = X_1X_2$. In this case, the PTx transmits two independent messages at a total rate $I(X;Y)$ and the RIS does not transmit messages. 
Compared with the case where $(X_1, X_2)$ are symbols transmitted at the PTx and RIS, respectively, the C-Rx receives the same messages. Thus, the above two transmission cases are equivalent for the C-Rx such that $I(X_1, X_2;Y) = I( X;Y)$. Therefore, maximizing $I(X_1, X_2;Y)$ is equivalent to maximizing $I(X;Y)$.

Due to $\mathbb{E}[|X_1|^2]\leq P$ and $|X_2|= 1$, the signal $X$ should satisfy the average power constraint, i.e, $\mathbb{E}[|X|^2] = \mathbb{E}[|X_1X_2|^2] \leq P$.
According to~\cite{cover1999elements}, under the average power constraint, when $ X \sim \mathcal{CN}(0,P)$ and $\varphi_k = e^{j(\vartheta-\mathrm{arg}(v_kg_k))}$, $I(X;Y)$ achieves the maximum, which is given by
\begin{equation}
 \label{eq:sum_new}
{C}_{\mathrm{sum}} = \log(1+\frac{P\tilde h^2}{\sigma^2}).
\end{equation}
This indicates that when $X_1X_2 \sim \mathcal{CN}(0,P)$, the RIS-assisted SR can achieve the maximum sum rate.
There are infinite distributions for $X_1$ and $X_2$ to achieve $X_1X_2 \sim \mathcal{CN}(0,P)$, which will be discussed in Section~\ref{sec:op}.

\begin{figure}[t]
\centering
\includegraphics[width=.85\columnwidth] {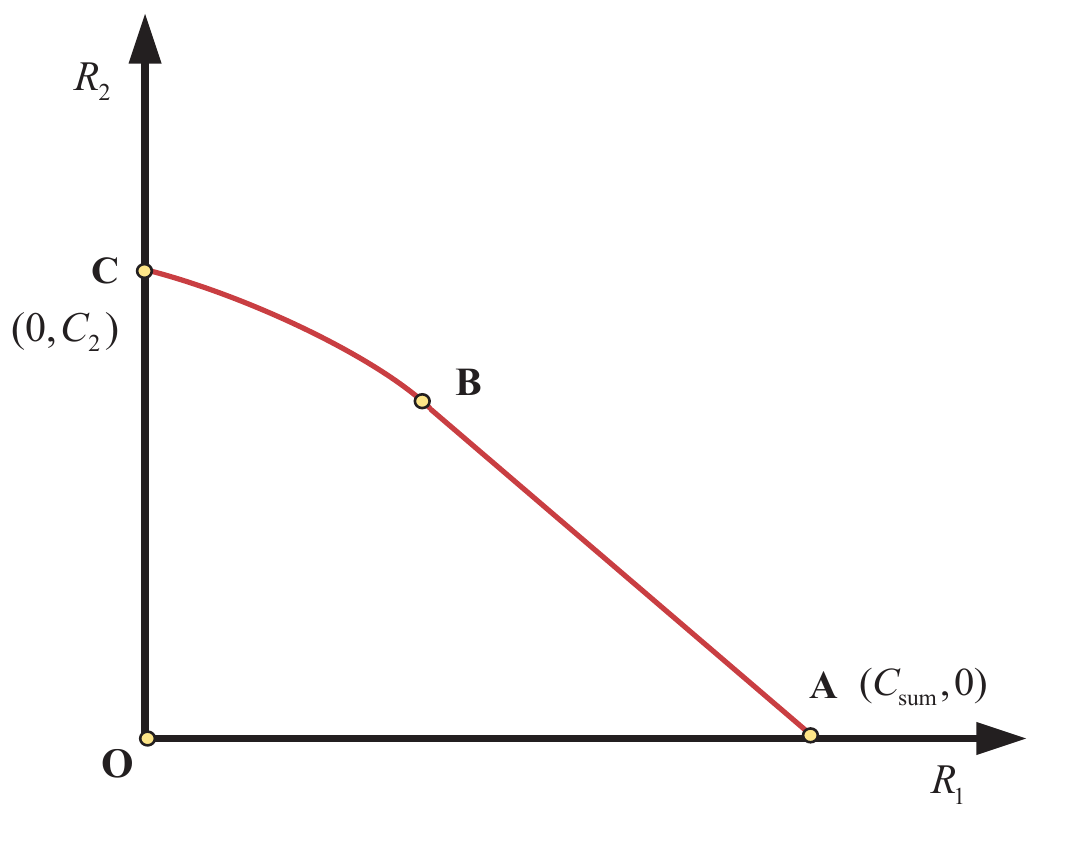}
\caption{Structure of capacity region of RIS-assisted SR.}
\label{fig:structure2}
\end{figure}

In the above subsections, we have characterized the maximum achievable rates of primary and secondary transmissions as
well as their maximum sum rate. We know that the maximum sum rate is equal to the maximum achievable rate of the primary transmission and is greater than the maximum achievable rate of the secondary transmission. Based on these features, we plot the capacity region structure of RIS-assisted SR, as shown in Fig.~\ref{fig:structure2}. Rate pairs on boundary ${\mathbf{A}}$-${\mathbf{B}}$ can achieve the maximum sum rate, which will be discussed in Section~\ref{sec:op}, while rate pairs on boundary ${\mathbf{B}}$-${\mathbf{C}}$ can not achieve the maximum sum rate
and present variations from the maximum sum rate to the maximum achievable rate of the secondary transmission, which will be discussed 
in Section~\ref{sec:op2}. 

\subsection{Rate Pairs on Boundary ${\mathbf{A}}$-${\mathbf{B}}$}\label{sec:op}

From Section~\ref{sec:Rsum}, the maximum sum rate is achievable when $X_1X_2 \sim \mathcal{CN}(0,P)$.
Since the amplitude of $X_2$ is equal to one, the amplitude of $X_1$ should follow a Rayleigh distribution to realize $X_1X_2 \sim \mathcal{CN}(0,P)$. There are infinite distributions for the phases of $X_1$ and $X_2$ to realize $X_1X_2 \sim \mathcal{CN}(0,P)$, which can be summarized as:\footnote{When $\theta_1\sim U[-\pi,\pi)$ and $\theta_2\sim U[-\pi,\pi)$, we also have $X\sim U[-\pi,\pi)$. However, 
in this case, the phase information of $X_1$ and $X_2$ is coupled together and may cause information ambiguity. Thus, in this paper, we do not consider this scenario.}
\begin{itemize}
  \item[(1)] Scheme I: The phase of $X_1$ follows a uniform distribution over $[-\alpha,\alpha)$, i.e., $\theta_1\sim U[-\alpha,\alpha)$. The phase of $X_2$ follows $\frac{\alpha}{\pi}\sum_n\delta(\theta_2-2n\alpha)$, where $n = 0,1,\cdots,\frac{\pi}{\alpha}-1$. Note that $\alpha$ needs to be carefully chosen such that $\frac{\pi}{\alpha}$ is a positive integer and $\alpha \neq 0$.
  \item[(2)] Scheme II: The phase of $X_2$ follows a uniform distribution over $[-\alpha,\alpha)$, i.e., $\theta_2\sim U[-\alpha,\alpha)$. The phase of $X_1$ follows $\frac{\alpha}{\pi}\sum_n\delta(\theta_1-2n\alpha)$, where $n = 0,\cdots,\frac{\pi}{\alpha}-1$, and  $\frac{\pi}{\alpha}$ is a positive integer with $\alpha \neq 0$.
\end{itemize}

\begin{figure}[t]
\centering
\includegraphics[width=.9\columnwidth] {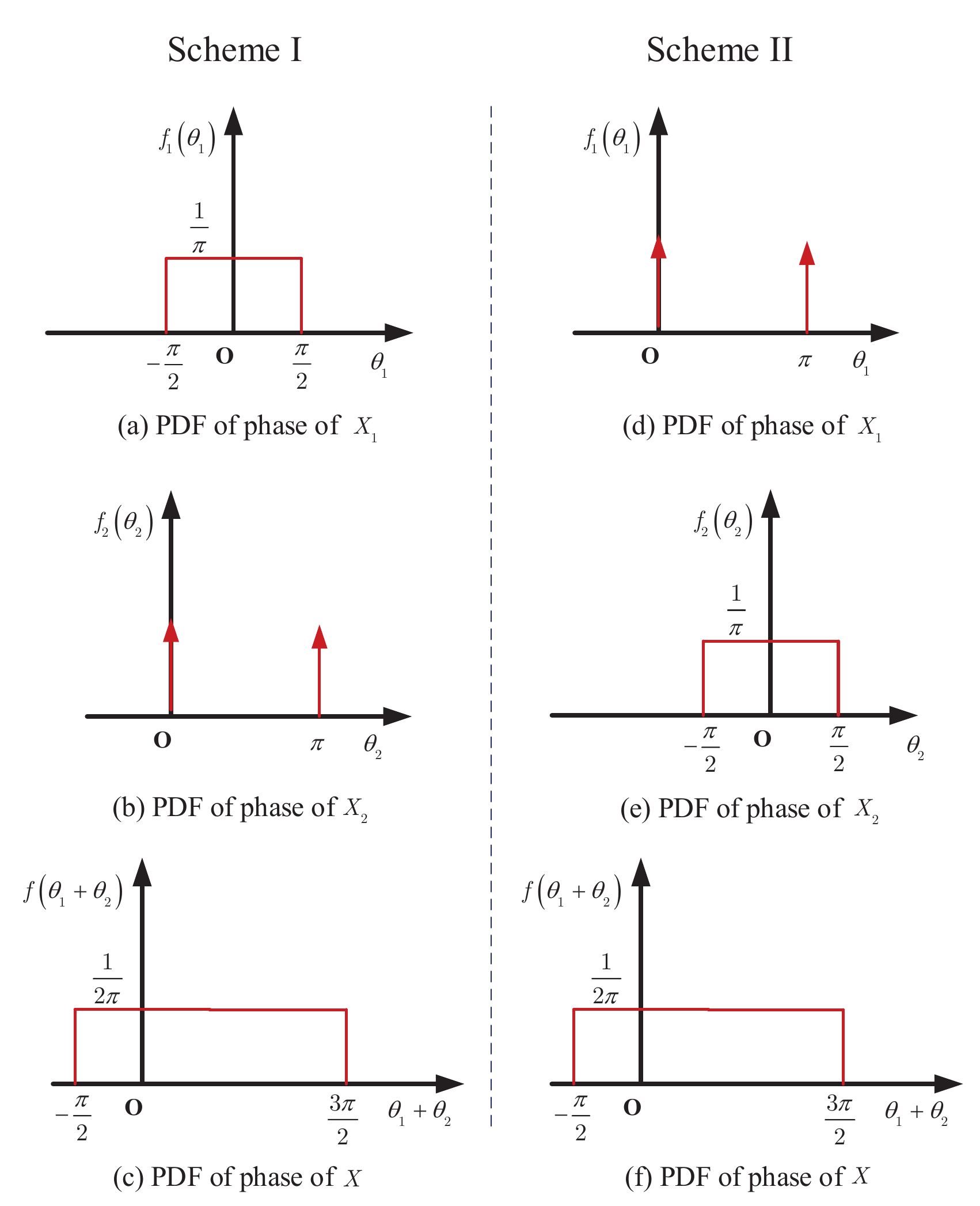}
\caption{Examples of the phase PDFs of $X_1$, $X_2$, and $X$.}
\label{fig:pdf}
\vspace{-1em}
\end{figure}

With scheme I and scheme II, the phase of $X_1X_2$ is uniformed distributed over $[-\pi, \pi)$.
Fig.~\ref{fig:pdf} shows examples of the phase PDFs of $X_1$, $X_2$, and $X$ for schemes I and II with $\alpha = \frac{\pi}{2}$.
By going through $\alpha$ from $0$ to $\pi$ and holding $\frac{\pi}{\alpha}$ be a positive integer for schemes I and II, we can obtain different rate pairs on boundary ${\mathbf{A}}$-${\mathbf{B}}$ with the maximum sum rate, which are given by Theorem~\ref{theorem:3}.
\begin{theorem}\label{theorem:3}
With the input distributions described in schemes I and II, the mutual information with a given decoding order can be calculated by
\begin{align*}
 &I(X_i;{Y})= \log(\pi e(P\tilde h^2+\sigma^2))+\mathbb{E}_{x_i}\int_{0}^{+\infty}\!\!\int_{-\pi}^{\pi}\kappa_{i,w}({r},{\psi},x_i)\\
&\qquad\qquad\times\log(\frac{\kappa_{i,w}({r},{\psi},x_i)}{r})d{\psi}  d{r},\\
&I(X_{t,t\neq i};{Y}|X_i)= - \mathbb{E}_{x_i}\int_{0}^{+\infty}\!\!\int_{-\pi}^{\pi}\kappa_{i,w}({r},{\psi},x_i)\\
&\qquad\qquad\times\log(\frac{\kappa_{i,w}({r},{\psi},x_i)}{r}) d{\psi}  d{r}-\log(\pi e\sigma^2),
\end{align*}
where $t,i\in\{1,2\}$, $w\in\{1,2\}$, and $\kappa_{i,w}({r},{\psi},x_i)$ for $i\in\{1,2\}$ and $w\in\{1,2\}$ are given by
\begin{itemize}
  \item[1)] when decoding $X_1$ first with scheme I, we have $\kappa_{1,1}({r},{\psi},x_1) = 
  \sum_{n=0}^{\frac{\pi}{\alpha}-1}\frac{{r}\alpha}{\pi^2\sigma^2}$
  $\exp\left(-\frac{{r}^2+\tilde h^2|x_1|^2-2{r}
  \tilde h|x_1|\cos(\theta_1+2n\alpha-{\psi})}{\sigma^2}\right)$;
  \item[2)] when decoding $X_2$ first with scheme I, we have $\kappa_{2,1}({r},{\psi},x_2)\!= \!\int_{0}^{+\infty} \!\int_{-\alpha}^{\alpha}\frac{{r}|x_1|}{2P\alpha\pi^2\sigma^2}$
  $\exp\left(-\frac{|x_1|^2}{P}-\frac{{r}^2+\tilde h^2|x_1|^2-2{r}\tilde h|x_1|
  \cos(\theta_1+\theta_2-\tilde{\psi})}{\sigma^2}\right)d\theta_1d|x_1|$;
  \item[3)] when decoding $X_1$ first with scheme II, we have $\kappa_{1,2}({r},{\psi},x_1) = \int_{-\alpha}^{\alpha}
  \frac{{r}}{2\alpha\pi\sigma^2}$
  $\exp\left(-\frac{{r}^2+\tilde h^2|x_1|^2-2{r}\tilde h|x_1|\cos(\theta_1+\theta_2-{\psi})}
  {\sigma^2}\right)d \theta_2$;
  \item[4)] when decoding $X_2$ first with scheme II, we have $\kappa_{2,2}({r},{\psi},x_2)  = \int_{0}^{+\infty}\sum_{n=0}^{\frac{\pi}{\alpha}-1}\frac{{r}|x_1|\alpha}{P\pi^3\sigma^2}$
$\exp\left(-\frac{|x_1|^2}{P}-\frac{{r}^2+\tilde h^2|x_1|^2-2{r}\tilde h|x_1|\cos(2n\alpha+\theta_2-{\psi})}{\sigma^2}\right)d|x_1|$.
\end{itemize}
\end{theorem}
\begin{IEEEproof}
Please see Appendix \ref{proof:2}.
\end{IEEEproof}

For scheme I, when $\alpha = \pi$, RIS does not transmit information and only assists the primary transmission by passive beamforming, i.e., $X_2 = 1$ and $X_1\sim \mathcal{CN}(0,P)$. For scheme II, when $\alpha = \pi$, the PTx only transmits the amplitude signal following Rayleigh distribution while the RIS transmits the phase signal following uniform distribution over $[-\pi,\pi)$. It should be noted that with $\alpha = \pi$ for scheme II, when $X_1$ is decoded first, $I(X_2;{Y}|X_1)$ achieves the maximum under the condition of the maximum sum rate.
Hence, the corresponding rate pair is the corner point ${\mathbf B}$ in Fig.~\ref{fig:structure2}.

The rate pairs characterized by Theorem~\ref{theorem:3} are some discrete points on boundary ${\mathbf{A}}$-${\mathbf{B}}$. The other rate pairs on boundary ${\mathbf{A}}$-${\mathbf{B}}$ can be achieved by time sharing. In this case, the cardinality of the time-sharing random variable is $|Q|\leq2$ according to~\cite[Appendix 4A]{el2011network}.

\vspace{-0.5em}
\subsection{Rate Pairs on Boundary ${\mathbf{B}}$-${\mathbf{C}}$}\label{sec:op2}

To characterize boundary ${\mathbf{B}}$-${\mathbf{C}}$, we need to further increase the rate of the secondary transmission.
Since the capacity region is the convex hull of rate pairs $(R_1,R_2)$, boundary ${\mathbf{B}}$-${\mathbf{C}}$ of the capacity region can be characterized by maximizing $\mu_1R_1+\mu_2R_2$ over all rate pairs, where $(\mu_1,\mu_2)$ are nonnegative and satisfy $\mu_1+\mu_2 = 1$ and $\mu_2>\mu_1$~\cite{goldsmith2003capacity}. Since the user with the lower priority will be decoded first, the maximization of $\mu_1R_1+\mu_2R_2$ can be written as $\mu_1I(X_1;{Y})+\mu_2I(X_2;{Y}|X_1)$~\cite{goldsmith2003capacity}.
Then, the rate pairs on boundary ${\mathbf{B}}$-${\mathbf{C}}$ can be obtained by solving the following optimization problem.
\begin{align*}
\textbf{P1}:  \;\; \mathop {\max}\limits_{\bm \varphi,f_1(X_1),f_2(X_2)}\;\; &\mu_1I(X_1;{Y})+\mu_2I(X_2;{Y}|X_1)\nonumber \\
\textrm{s.t.}~\;\; \qquad& \eqref{eq:c1}, \eqref{eq:c2}, \textrm{and }\eqref{eq:c3}.
\end{align*}

Since the stronger channel response enables higher achievable rates of primary and secondary transmissions, $\bm \varphi$ can be designed to maximize $|h|^2$, whose solution is $\varphi_k = e^{j(\vartheta-\mathrm{arg}(v_kg_k))}$. 
Then, we focus on the optimal distributions of $X_1$ and $X_2$.
The objective function can be written as 
 \begin{align*}
&\mu_1I(X_1;{Y})+\mu_2I(X_2;{Y}|X_1)\\
=&\mu_1H(Y)-\mu_1H(Y|X_1)+\mu_2H(Y|X_1)-\mu_2H(Y|X_1,X_2)\\
=&\mu_1H(Y)+(\mu_2-\mu_1)H(Y|X_1)-\mu_2H(Y|X_1,X_2).
\end{align*}

The maximization of $H(Y)$ requires $\theta_1+\theta_2 \sim U[ -\pi,\pi)$, while the maximization of $H(Y|X_1)$ requires $\theta_2 \sim U[ -\pi,\pi)$, which has been discussed in Sections~\ref{sec:R1} and~\ref{sec:R2}. Due to $\mu_2-\mu_1>0$, it is obvious that the optimal distribution of $f_2(X_2)$ is characterized by a uniformly distributed independent phase from $-\pi$ to $\pi$, while the PTx only transmits amplitude information. Accordingly, the PDF of $X_1$ is given by $f_1(X_1) = f_a(A)\delta(\theta_1)$, where $A$ is the amplitude of $X_1$ and $f_a(A)$ is the PDF of $A$.
Then, the optimization problem $\textbf{P1}$ can be recast as
\begin{align*}
\textbf{P2}:  \;\; \mathop {\max}\limits_{f_a}\;\; &\mu_1H(Y)+(\mu_2-\mu_1)H(Y|A)\nonumber \\
\textrm{s.t.}~\;\; & \int_{0}^{\infty}f_a a^2 d a\leq P.
\end{align*}

Here, we introduce the marginal entropy induced by $f_a$, which is given by
\begin{align}
\omega(A;f_a) = -\int f_y(Y|A)\log f_y(Y;f_a)dy,
\end{align}
where $f_y(Y|A)$ is the PDF of $Y$ given $A$ and $f_y(Y;f_a)$ is the PDF of $Y$ with $f_a$.
The following theorem provides the properties of the optimal $f_a$ in problem $\textbf{P2}$.

\begin{theorem}\label{theorem:feature}
The optimal distribution of $A$ is denoted by $f_a^*$. Then, $f_a^*$ is optimal if and only if
\begin{align}
&\omega(A;f_a^{*})\!\leq\! T_0-\left(\frac{\mu_2}{\mu_1}-1\right)H(Y|A=a)+\frac{\lambda}{\mu_1} a^2,~\forall a \label{eq:condtion1}\\
&\omega(A;f_a^{*})\!=\! T_0-\left(\frac{\mu_2}{\mu_1}-1\right)H(Y|A=a)+\frac{\lambda}{\mu_1} a^2, \forall a \in E_0,\label{eq:condtion2}
\end{align}
where $\lambda \geq 0$ is a Lagrange multiplier, $T_0$ is a constant value and equal to $\int_0^{\infty} f_a^*\omega(A;f_a^{*})da+(\frac{\mu_2}{\mu_1}-1)\int_0^{\infty} f_a^*H(Y|A=a)da-\frac{\lambda}{\mu_1} P$, and $E_0$ is the points of increase of the cumulative distribution function (CDF) $F_a^*$ that corresponds to $f_a^{*}$.
\end{theorem}
\begin{IEEEproof}
Please see Appendix~\ref{proof:feature}.
\end{IEEEproof}

From Appendix~\ref{proof:finite}, the increase points set $E_0$ of the optimal CDF $F_a^*$ is finite. Thus, the optimal PDF $f_a^*$ can be written as 
\begin{align}\label{eq:pdfgeneral}
f_a^*(A)=\sum_{m = 1}^M {{p_m }} \delta (A - {a_m}), 
\end{align}
where $\sum_{m=1}^{M}p_m{a_m^2} \leq P$, $a_{m+1}>a_m$, $0\leq p_m\leq 1$, and $\sum_{m = 1}^{M}p_m  = 1$.
Meanwhile, $M$ is the number of mass points, $a_m$ is the location of the $m$-th mass point, and $p_m$ is the corresponding probability of the $m$-th mass point. Note that the number, location, and probability of mass points in~\eqref{eq:pdfgeneral} should satisfy the conditions in
Theorem~\ref{theorem:feature}.

With $f_a^*(A)$ in~\eqref{eq:pdfgeneral}, according to Appendix~\ref{proof:1}, the achievable rates for both primary and secondary transmissions are given, respectively, by
\begin{align}
&I(X_1;{Y})= \sum_{m= 1}^{M}p_m\int_{0}^{+\infty}\frac{2{r}\bar{\kappa}({r},a_m)}{\sigma^2} \log\left({\bar{\kappa}({r},a_m)}\right) d {r}\nonumber\\
&\!-\!\int_{0}^{+\infty}\!\!\frac{\sum_{m = 1}^{M}2{r}p_m\bar{\kappa}({r},a_m)}{\sigma^2}\! \log\!\left(\!\sum_{m = 1}^{M}p_m\bar{\kappa}({r},a_m)\!\!\right)\! d{r},\label{eq:R1}\\
&I(X_2;{Y}|X_1)= -\!\sum_{m= 1}^{M}p_m\int_{0}^{+\infty}\frac{2{r}\bar{\kappa}({r},a_m)}{\sigma^2} \nonumber\\ &\:\qquad\qquad\qquad\qquad\times\log\left({\bar{\kappa}({r},a_m)}\right) d{r}-\log(e),\label{eq:R2}
\end{align}
where $\bar{\kappa}({r},\!a_m) =\exp\!\left(\!-\frac{{r}^2+\tilde h^2a_m^2}{\sigma^2}\!\right)\!I_0\!\left(\!\frac{2{r}\tilde ha_m}{\sigma^2}\!\right)$.

Although we prove that the optimal distribution of the amplitude of $X_1$ of rate pairs on boundary $\mathbf B$-$\mathbf C$ is discrete, a 
closed-form solution of $f_a^*$ with specific $M$, $p_m$, and $a_m$ seems unlikely. On one hand, the optimal $M$, $p_m$, and $a_m$ can be
obtained by using the scheme proposed in~\cite{smith1971information}, which solves $f_a^*$ based on the conditions in~\eqref{eq:condtion1}
and~\eqref{eq:condtion2}. On the other hand, the optimal $M$, $p_m$, and $a_m$ can be obtained by solving the following optimization problem
directly:
\begin{align}
\textbf{P3}:  \;\;\mathop {\max}\limits_{M,a_m,p_m}\;\; &\mu_1I(X_1;{Y})+\mu_2I(X_2;{Y}|X_1) \nonumber \\
\textrm{s.t.}~~~~~~\;\;& \sum_{m=1}^{M}p_m{a_m^2} \leq P, ~~ a_{m+1}> a_m\nonumber \\
&0\leq p_m\leq 1, ~~ \sum_{m = 1}^{M}p_m  = 1.\nonumber
\end{align}
Note that the objective function in problem $\textbf{P3}$ can be explicitly represented by~\eqref{eq:R1} and~\eqref{eq:R2}.
For problem $\textbf{P3}$, the variables $M$, $a_m$, and $p_m$ are coupled together in the objective function. Thus,
the objective function is not concave over $M$, $a_m$, and $p_m$.
Meanwhile, one can find that the constraint $\sum_{m=1}^{M}p_m{a_m^2} = P$ is not a convex set. Thus, problem $\textbf{P3}$ is a 
non-convex optimization problem and belongs to nonlinear programming.
Nonetheless, it can be solved by using an interior-point algorithm, which introduces barrier functions to deal with inequality 
constraints and adopts conjugate gradient methods to find a convergence point~\cite{nocedal2014interior}. Due to space limitations, 
we omit the details of interior-point algorithms.

\section{Capacity Region Characterization with Constraint $|X_2|\leq1$}\label{sec:Rate2}

In this section, we will characterize the capacity region with constraint $|X_2|\leq1$.
Following the derivations in Section~\ref{sec:Rate}, one can find that the adjustment of the amplitude will not affect the maximum achievable rate of the primary transmission and the maximum sum rate. Thus, in the following, we will focus on the maximum achievable rate of the secondary transmission and the rate pairs on boundary ${\mathbf{B}}$-${\mathbf{C}}$.

\subsection{Maximum Achievable Rate of Secondary Transmission}\label{sec:R22}

\begin{figure}[t]
\centering
\includegraphics[width=.6\columnwidth] {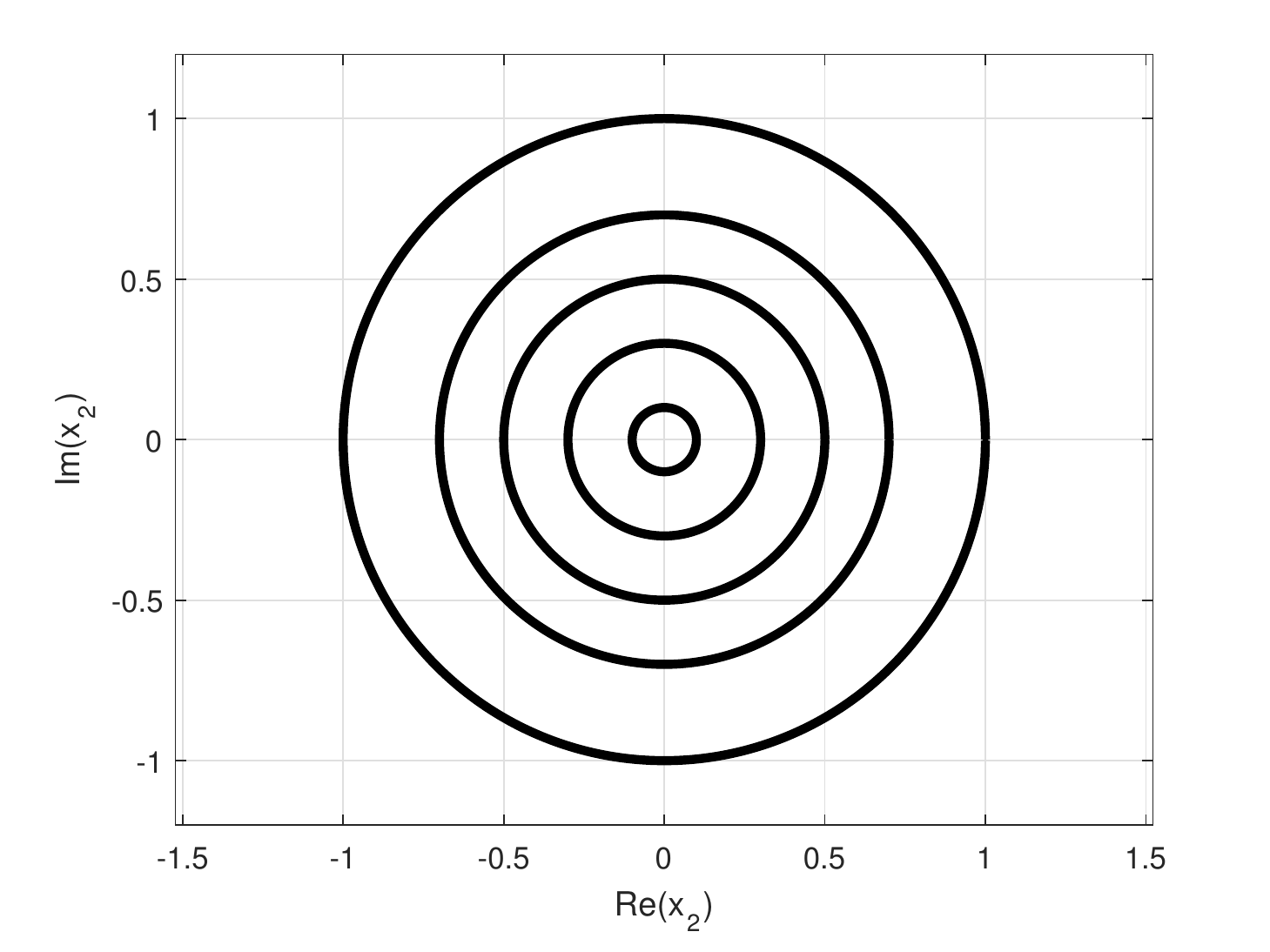}
\caption{An example of the optimal input distribution with passive reflection constraint.}
\label{fig:distribution}
\vspace{-1em}
\end{figure}

Mathematically, the capacity under passive reflection constraint $|X_2|\leq 1$ is equivalent to that under a peak power constraint $|X_2|^2\leq 1$. Thus, the optimal distribution to maximize $I(X_2;Y|X_1 = x_1)$ is geometrically characterized by concentric circles, i.e., discrete amplitude and uniform independent phase, as shown in Fig.~\ref{fig:distribution}~\cite{shamai1995capacity}.
Denote the amplitude of $X_2$ by $r$, whose PDF $f_{r}$ is given by
\begin{equation}\label{eq:pdf1}
f_{r}(r) = \sum_{\ell = 1}^{L}q_\ell \delta(r-r_l),
\end{equation}
with $0\leq r_\ell\leq 1$, $r_{\ell+1}> r_\ell$, $0\leq q_\ell\leq 1$, and $\sum_{\ell = 1}^{L}q_\ell  = 1$. Meanwhile, $L$ is the number of mass points, $r_l$ is the location of the $\ell$-th mass point, and $q_\ell$ is the corresponding probability of the $\ell$-th mass point.
Note that the number of mass points and the corresponding locations and probabilities are determined by maximizing the mutual information of the secondary transmission. 

Following the derivations in Appendix~\ref{proof:capacity}, we know that under constraint $|X_1|\leq 1$, the mutual information $I(X_2;Y|X_1 = x_1)$ has the same expression with~\eqref{eq:sec1} but with $\kappa(r) = \sum_{\ell = 1}^{L}q_\ell \exp\left(-\frac{{r}^2+|hx_1|^2r_\ell^2}{\sigma^2}\right)I_0\left(\frac{2{r}|hx_1|r_\ell}{\sigma^2}\right)$.
Considering the complexity of the expression of $I(X_2;Y|X_1 = x_1)$, we provide an upper bound on $I(X_2;Y|X_1 = x_1)$, which is given by~\cite{thangaraj2017capacity}
\begin{align}\label{eq:upper2}
I(X_2;Y&|X_1 = x_1) \leq \min\left\{\log\left(1+\frac{\tilde{h}^2|x_1|^2}{\sigma^2}\right),\right.\nonumber\\
&~~~\left.\log\left(1+\sqrt{\frac{\pi \tilde{h}^2|x_1|^2}{\sigma^2}}+\frac{\tilde{h}^2|x_1|}{e\sigma^2}\right)\right\}.
\end{align}
In~\eqref{eq:upper2}, the first part in the min function represents the capacity relaxed to an average power constraint and the second part represents the McKellips-type bound. At low $\frac{\tilde{h}^2|x_1|^2}{\sigma^2}$, $I(X_2;Y|X_1 = x_1)$ approaches the first part in~\eqref{eq:upper2}, while approaches the second part at high $\frac{\tilde{h}^2|x_1|^2}{\sigma^2}$.

Following the derivations in Section~\ref{sec:R1}, we know that the optimal passive beamforming of RIS is $\varphi_k = e^{j(\vartheta-\mathrm{arg}(v_kg_k))}$ and the primary transmission satisfies $|X_1|^2 = P$ to maximize the achievable rate of the secondary transmission. Then, the maximum achievable rate of the secondary transmission under constraint $|X_2|\leq 1$ is given by
\begin{equation}
 \label{eq:secondary}
C_2 = \max_{L,p_l,r_l} \;\; -\int_{0}^{+\infty}\frac{2{r}\tilde{\kappa}({r})}{\sigma^2} \log\left(\tilde{\kappa}({r})\right) d{r}-\log(e),
\end{equation}
where $\tilde{\kappa}(r) = \sum_{\ell = 1}^{L}q_\ell \exp\left(-\frac{{r}^2+{P}\tilde h^2r_\ell^2}{\sigma^2}\right)$ $I_0\left(\frac{2{r}\sqrt{P}\tilde hr_\ell}{\sigma^2}\right)$.

When the receive SNR $\frac{P\tilde{h}^2}{\sigma^2}$ is low, the capacity-achieving distribution is geometrically characterized by a unit circle, i.e., the amplitude of $X_2$ is equal to $1$. With the growth of $\frac{P\tilde{h}^2}{\sigma^2}$, the concentric circle number of the capacity-achieving distribution will increase. An upper bound on~\eqref{eq:secondary} is given by~\cite{thangaraj2017capacity}
\begin{align}\label{eq:upper}
C_2 \leq \min&\left\{\log\left(1+\sqrt{\frac{\pi P\tilde{h}^2}{\sigma^2}}+\frac{P\tilde{h}^2}{e\sigma^2}\right)\right. ,\nonumber\\
&~~~\left.\log\left(1+\frac{P\tilde{h}^2}{\sigma^2}\right)\right\}.
\end{align}

It can be calculated that there is an SNR gap of about $10\log_{10}(e)\approx 4.34 $ dB at high receive SNR between the two parts in~\eqref{eq:upper}.
This indicates that if we use $\log\left(1+\frac{P\tilde{h}^2}{\sigma^2}\right)$ to analyze the performance of the secondary transmission, we will get a more than $4.34 $ dB performance gap at high receive SNR.

In addition, we can find that under constraint $|X_1| \leq 1$, the high SNR slope is $\lim_{P \rightarrow \infty} \frac{C_2}{\log P} = 1$, while under constraint $|X_1| = 1$, the high SNR slope becomes $\lim_{P \rightarrow \infty} \frac{C_2}{\log P} = \frac{1}{2}$.
This indicates that if RIS only transmits the phase information, the degree-of-freedom of the secondary transmission will be halved.


\begin{figure*}
  \begin{tabular}{cccccc}
   &\includegraphics[width=.6\columnwidth]{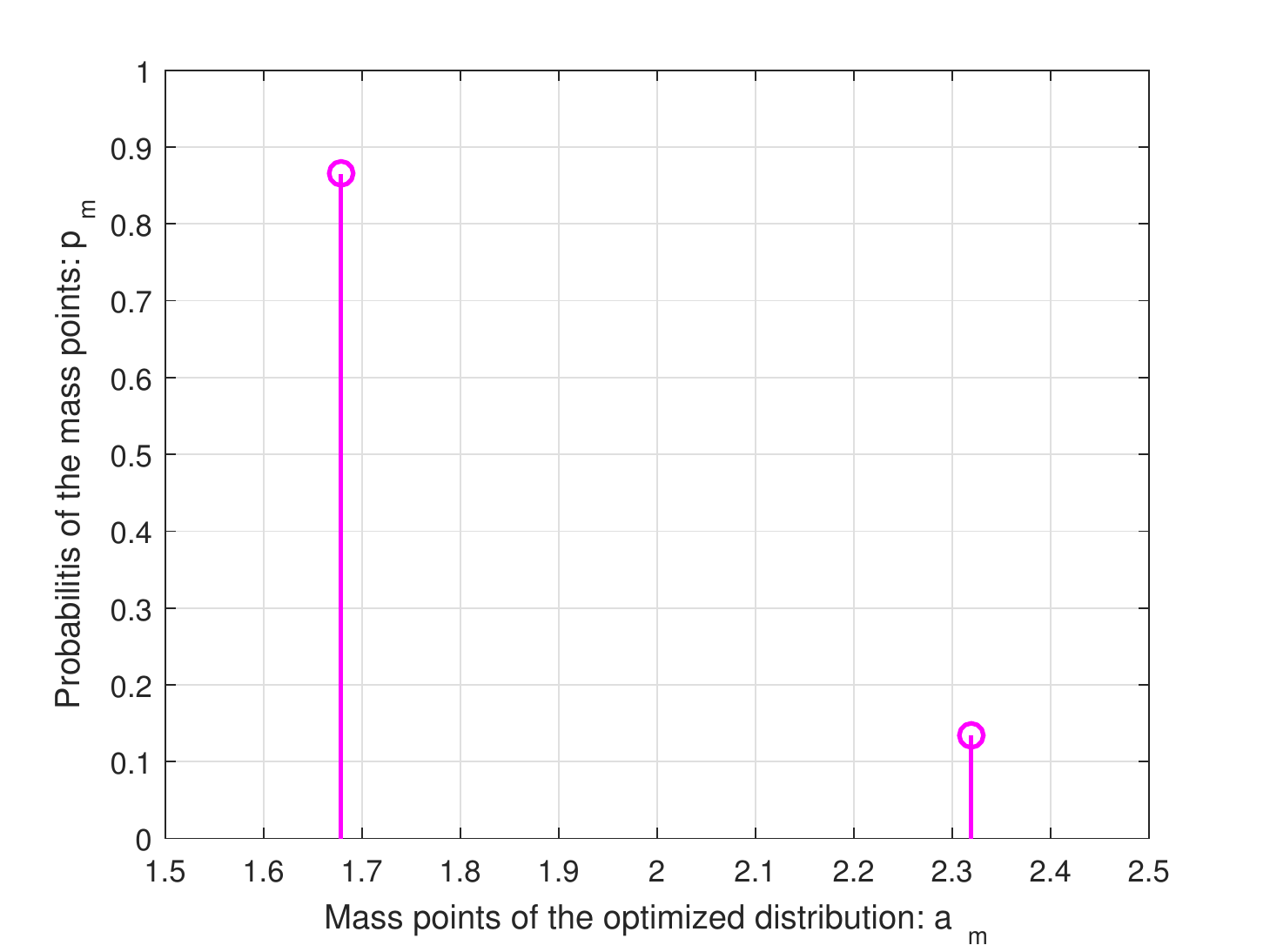} &\includegraphics[width=.6\columnwidth]{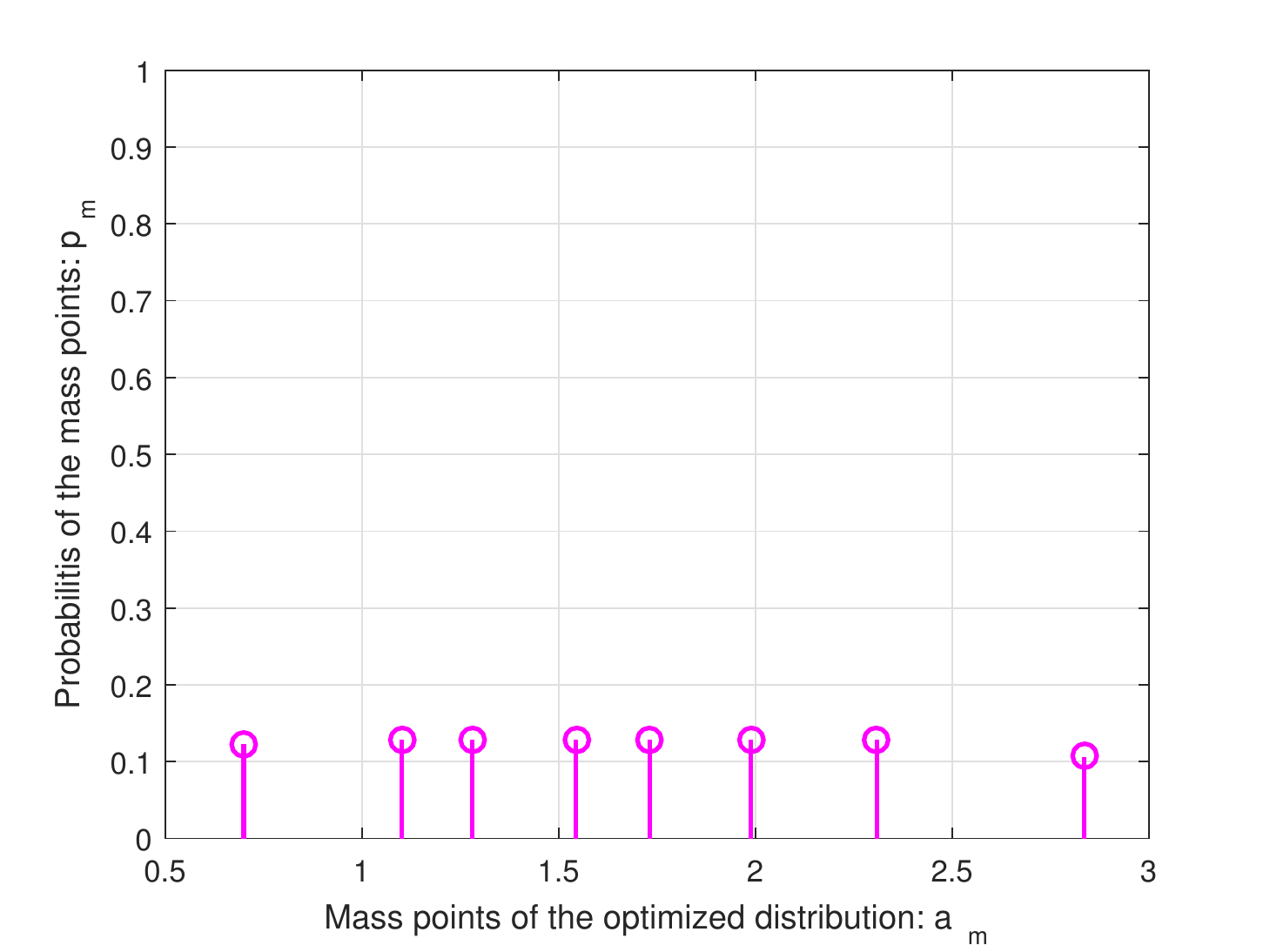} & \includegraphics[width=.6
   \columnwidth]{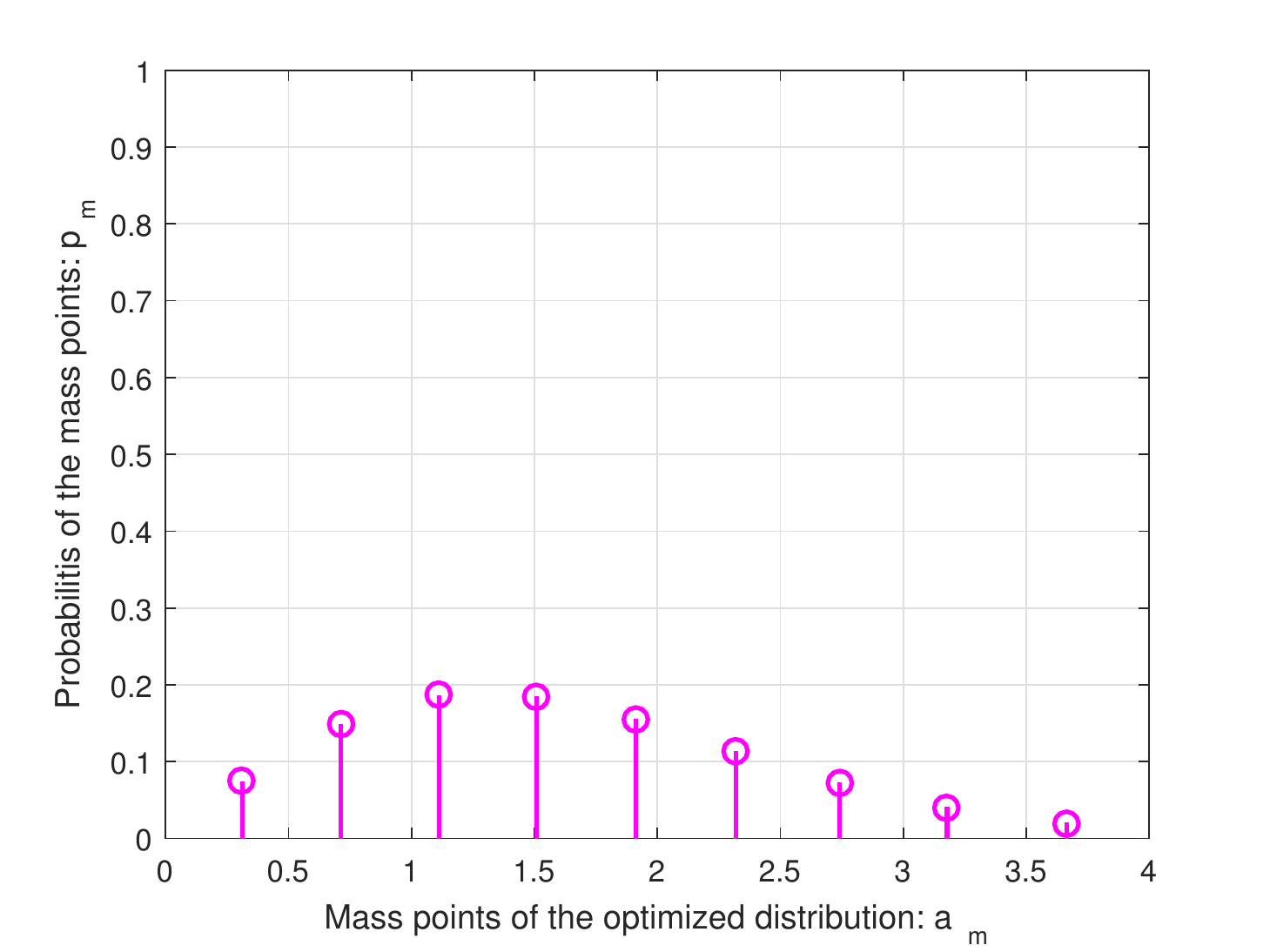} \\
   &(a)  $\mu_1 = 0.1$     &(b)  $\mu_1 = 0.3$    &(c)  $\mu_1 = 0.49$
  \end{tabular}
  \caption{The optimized locations of the mass points versus the corresponding probabilities
  of $f_a^*$ with constraint $|X_2| = 1$.}\label{fig:distribution1}
  \vspace{-1em}
\end{figure*}

\subsection{Rate Pairs on Boundary ${\mathbf{B}}$-${\mathbf{C}}$}\label{sec:op3}

With constraint $|X_1|\leq 1$, the rate pairs on boundary ${\mathbf{B}}$-${\mathbf{C}}$ can also be characterized by solving $\textbf{P1}$.
It is obvious that the design of the passive beamforming vector is the same as the case with $|X_1|=1$ such that the optimization problem $\textbf{P1}$ can be written as
\begin{align*}
\textbf{P4}:  \;\; \mathop {\max}\limits_{f_1(X_1),f_2(X_2)}\;\; &\mu_1H(Y)+(\mu_2-\mu_1)H(Y|X_1)\nonumber \\
\textrm{s.t.}\;\; \qquad& \eqref{eq:c1} ~\textrm{and } |X_1|\leq 1.
\end{align*}
With the same reason as the case $|X_2| = 1$, $|X_2|$ is characterized by a uniformly distributed independent phase from $-\pi$ to $\pi$, while the PTx only transmits amplitude information. 

From Appendix~\ref{proof:finite2}, the optimal amplitude PDFs of both $X_1$ and $X_2$ are discrete and given by~\eqref{eq:pdfgeneral}
 and~\eqref{eq:pdf1}, respectively. 
Thus, we can derive the mutual information $I(X_1;{Y})$ and $I(X_2;{Y}|X_1)$, which have the same expressions with~\eqref{eq:R1} 
and~\eqref{eq:R2}, respectively, but with $\bar{\kappa}({r},a_m) = \sum_{\ell = 1}^{L}q_{\ell} \exp\left(-\frac{{r}^2+\tilde h_2^2a_m^2r_\ell^2}{\sigma^2}\right)I_0\left(\frac{2{r}\tilde h_2a_mr_\ell}{\sigma^2}\right)$.

Note that the number, location, and probability of mass points in~\eqref{eq:pdfgeneral} and~\eqref{eq:pdf1} can be obtained by solving the following optimization problem:
\begin{align}
\textbf{P5}:  \;\;\mathop {\max}\limits_{L,M,r_\ell,a_m,q_\ell,p_m}\;\; &\mu_1I(X_1;{Y})+\mu_2I(X_2;{Y}|X_1) \nonumber \\
\textrm{s.t.}~~~~~~\;\;&  0\leq r_l\leq 1,~~  r_{\ell+1}> r_l, \nonumber \\
& 0\leq q_l\leq 1, ~~ \sum_{\ell = 1}^{L}q_\ell  = 1,\nonumber \\
& \sum_{m=1}^{M}p_m{a_m^2} \leq P, ~~ a_{m+1}> a_m\nonumber \\
&0\leq p_m\leq 1, ~~ \sum_{m = 1}^{M}p_m  = 1.\nonumber
\end{align}
Similar to problem $\textbf{P3}$, problem $\textbf{P5}$ can be solved by using an interior-point algorithm~\cite{nocedal2014interior}.

\begin{figure}
  \begin{tabular}{cccccc}
   &\includegraphics[width=.45\columnwidth]{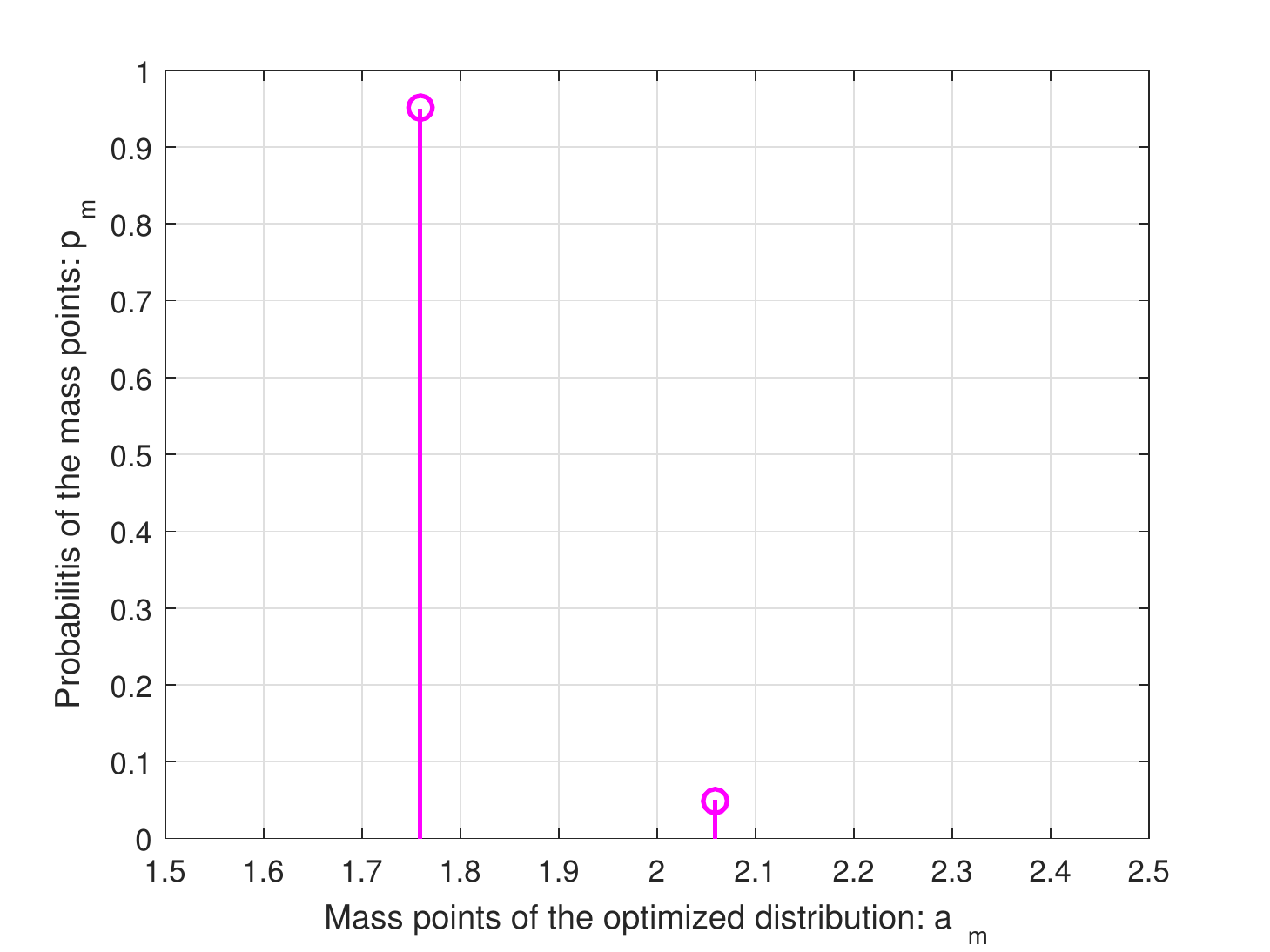}  &\includegraphics[width=.45\columnwidth]{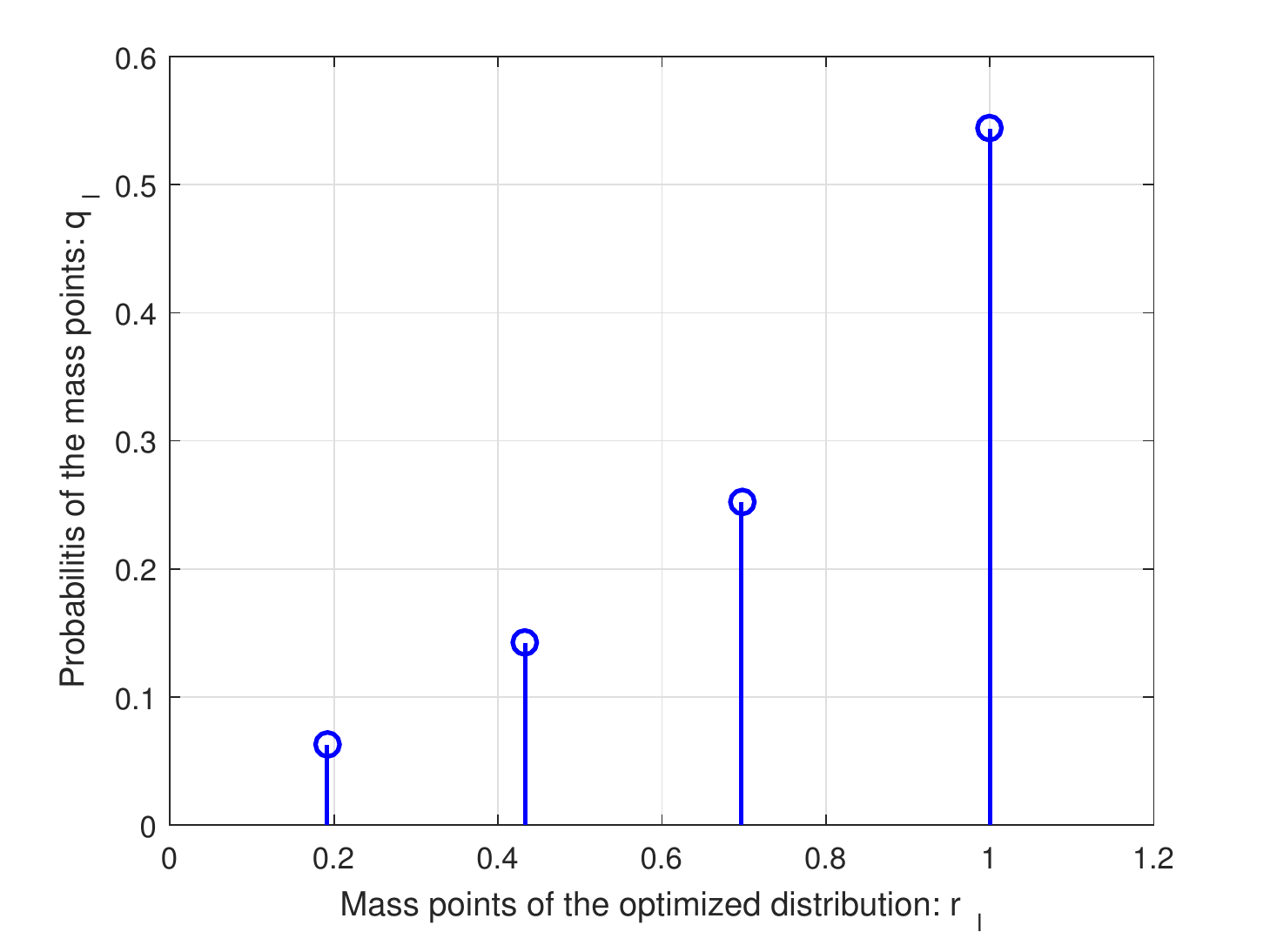} \\
   &(a) Primary: $\mu_1 = 0.1$ &(b) Secondary: $\mu_2 = 0.9$ \\
      &\includegraphics[width=.45\columnwidth]{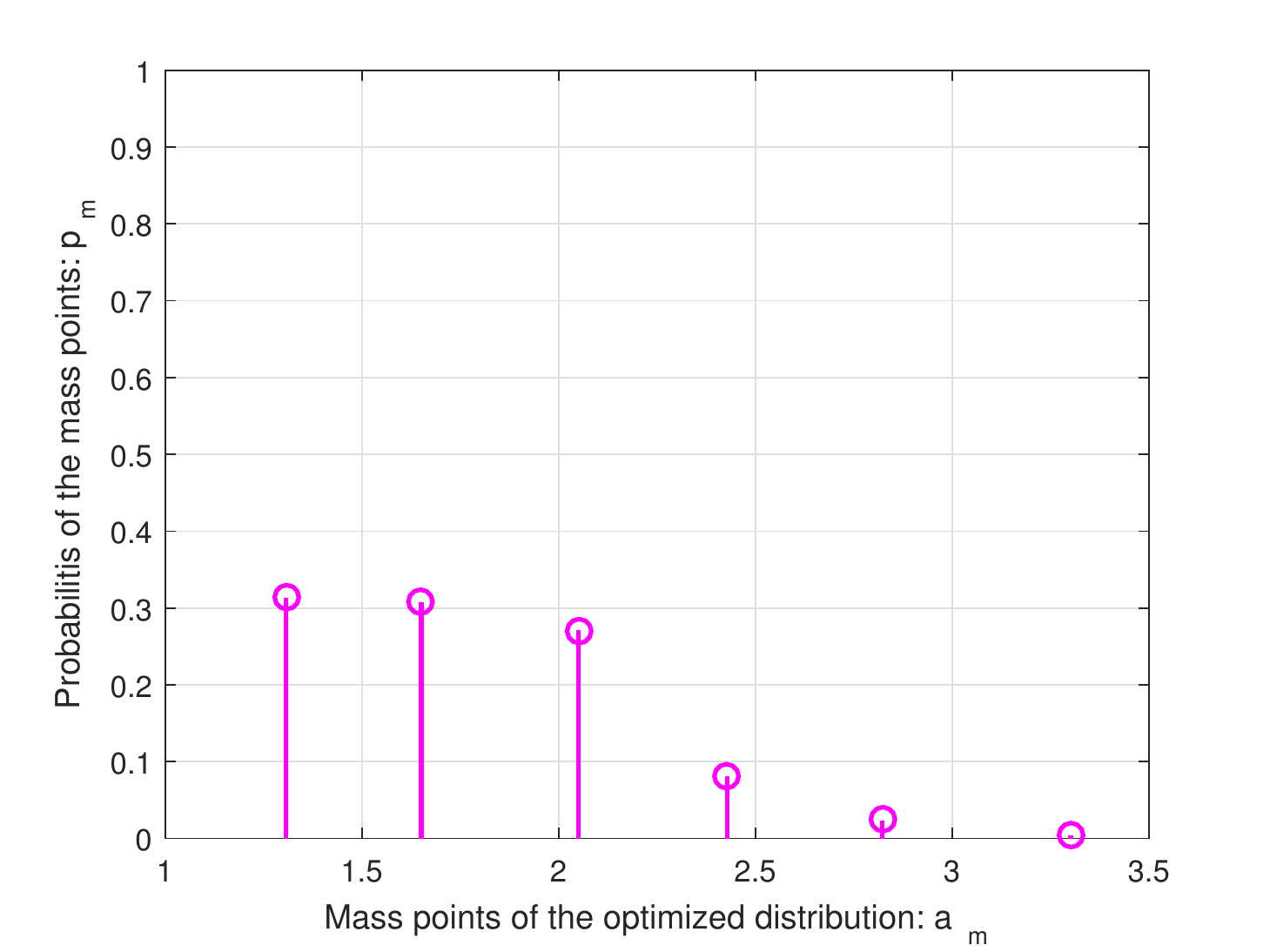}
   &\includegraphics[width=.45\columnwidth]{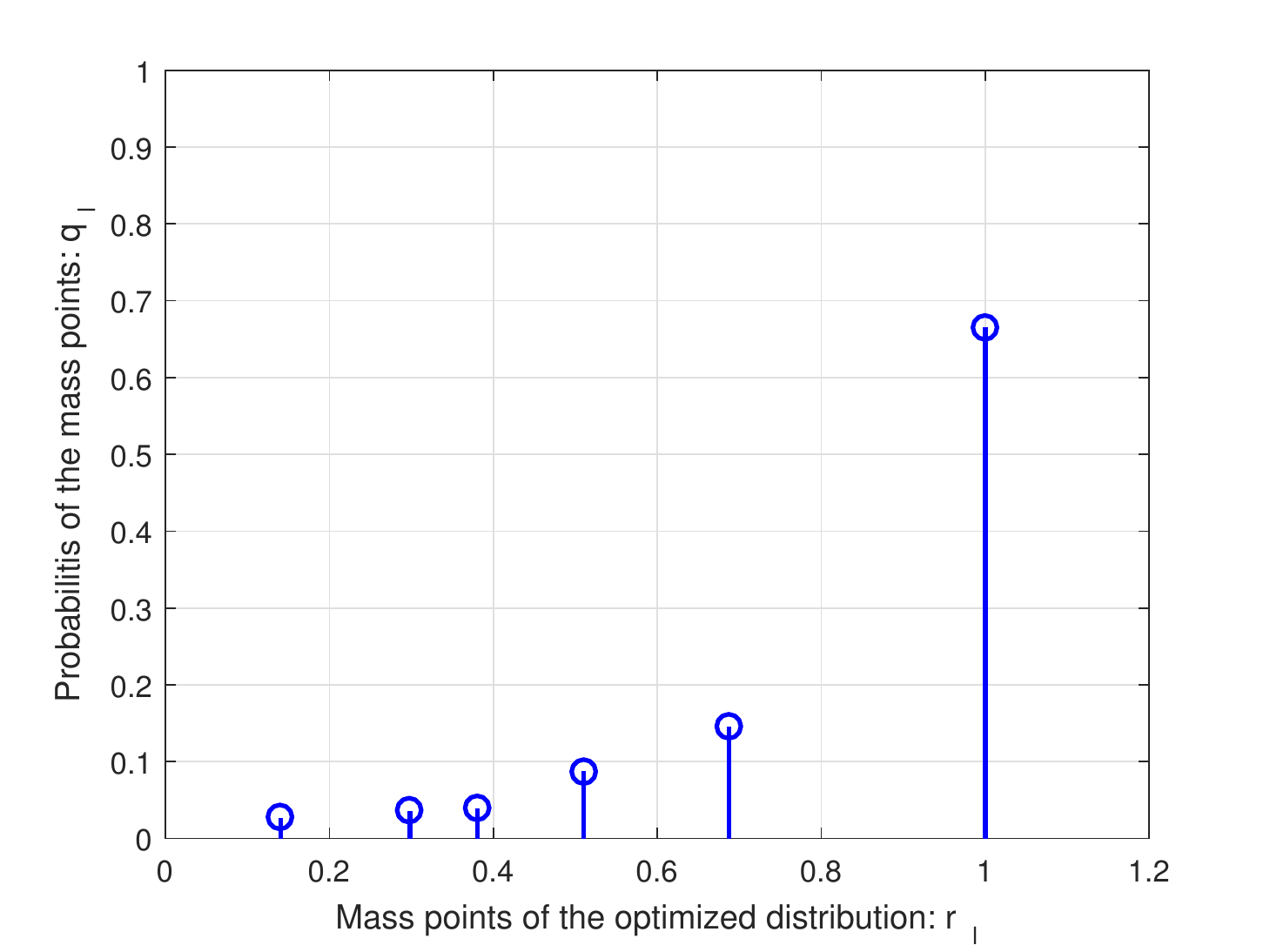} \\
    &(c) Primary: $\mu_1 = 0.3$   &(d) Secondary: $\mu_2 = 0.7$  \\
   &\includegraphics[width=.45\columnwidth]{49constant2.pdf} &\includegraphics[width=.45\columnwidth]{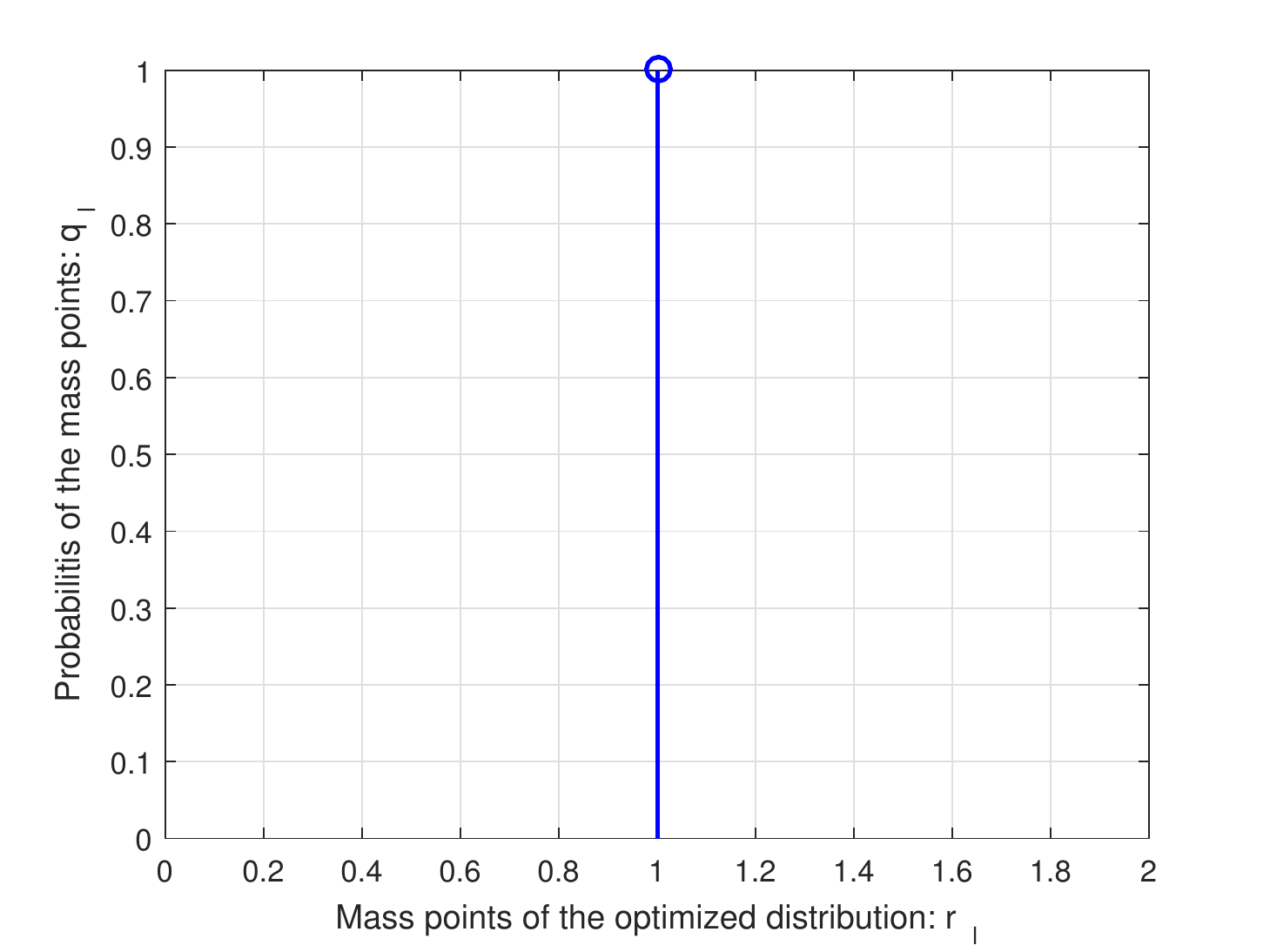}\\
   &(e) Primary: $\mu_1 = 0.49$ & (f) Secondary: $\mu_2 = 0.51$
  \end{tabular}
  \caption{The optimized locations of the mass points versus the corresponding probabilities for the distributions
  of the primary and secondary transmissions with constraint $|X_2| \leq 1$.}\label{fig:distribution2}
  \vspace{-1.5em}
\end{figure}

\section{Performance Evaluation} \label{sec:simulation}

In this section, numerical results are presented to evaluate the capacity region of RIS-assisted SR. 
Since the stronger reflecting link represents the higher receive SNR, the effect of $K$ on the performance of primary and secondary transmissions
is equivalent to the effect of $\frac{P}{\sigma^2}$. Thus, unless specified otherwise, we set $K = 64$ and $\rho^2|g_kv_k|^2 = 0.003$, for $k = 1,\cdots,K$.
Problems $\textbf{P3}$ and $\textbf{P5}$ are solved by using the optimization package of Matlab~\cite{MathWorks}.

Fig.~\ref{fig:distribution1} plots the optimized locations of the mass points versus the corresponding probabilities of $f_a^*$ with 
constraint $|X_2| = 1$. In this figure, we set $P/\sigma^2 = 5$ dB. We can find that 
when $|X_2| = 1$, the number of mass points increases with the growth of $\mu_1$, and the probability of $f_a^*$ approaches a Rayleigh distribution when $\mu_1 = 0.49$.
Fig.~\ref{fig:distribution2} plots the optimized locations of the mass points versus the corresponding probabilities of the distributions
for both the primary and secondary transmissions with constraint $|X_2|\leq 1$. We set $P/\sigma^2 = 5$ dB. 
Similar to Fig.~\ref{fig:distribution1}, with the increase of $\mu_1$, the number of mass points of $f_a^*$ increases.
When $\mu_2 = 0.51$, the optimal $X_2$ is equal to $X_2 = e^{j\theta_2}$ and the optimal $f_1^*$ approaches
a Rayleigh distribution. This indicates that the rate pair when $\mu_2 = 0.51$ approaches point ${\mathbf{B}}$ on the region boundary.

\begin{figure}[t]
\centering
\includegraphics[width=.9\columnwidth] {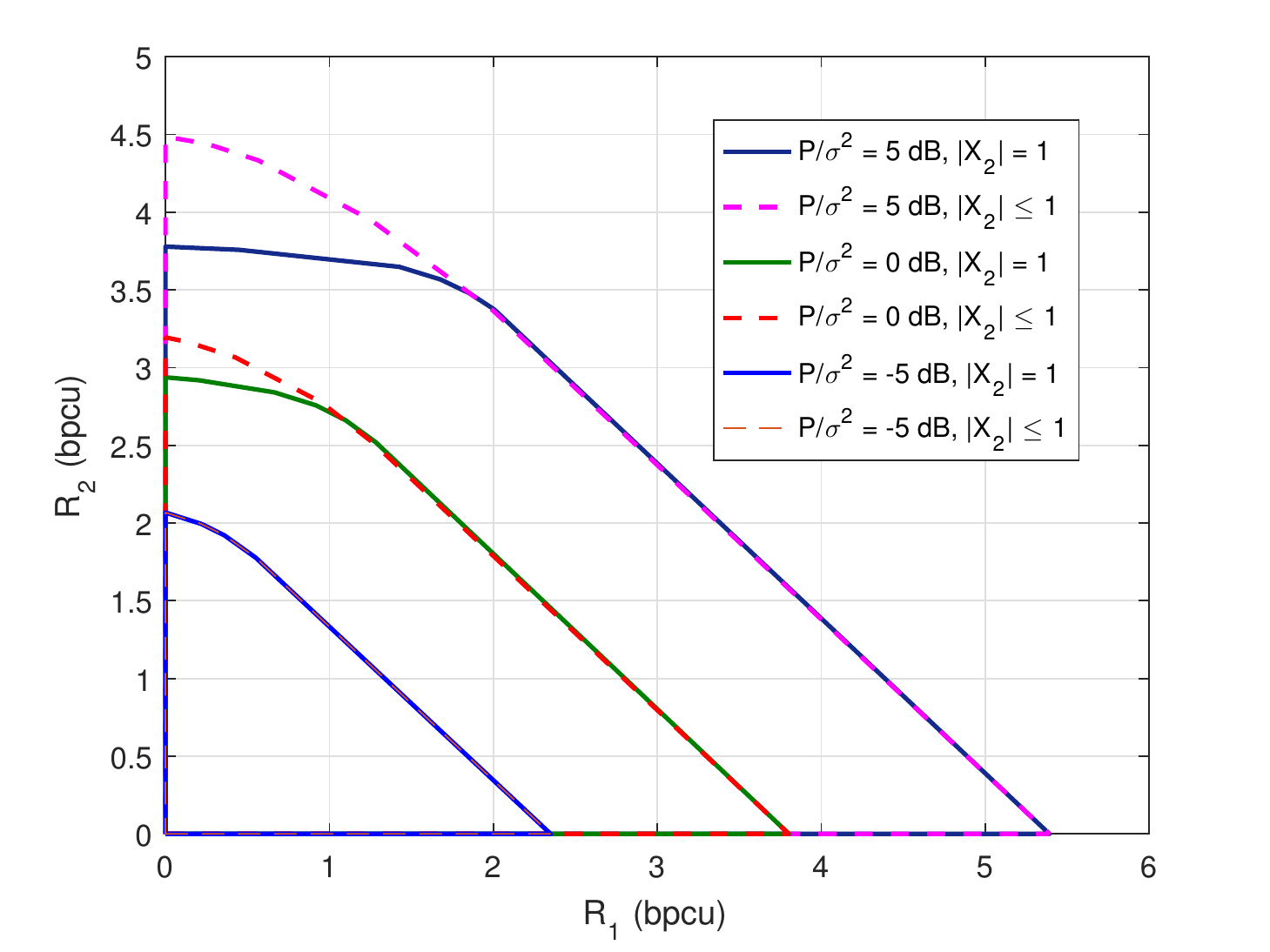}
\caption{Capacity region.}
\label{fig:region1}
\vspace{-1.5em}
\end{figure}

In Fig.~\ref{fig:region1}, the capacity region of RIS-assisted SR is plotted for different $P/\sigma^2$ with constraints $|X_2| = 1$ and $|X_1| \leq 1$. 
We can see that the capacity region is always strictly convex. A higher average transmit power at the PTx can lead to a 
larger capacity region. 
It is obvious that constraint $|X_2|\leq 1$ enables a larger capacity region for RIS-assisted SR compared with constraint $|X_2| = 1$
when $\frac{P}{\sigma^2} = 0$ dB and $\frac{P}{\sigma^2} = 5$ dB.
In addition, with the growth of $\frac{P}{\sigma^2}$, the gap between the capacity regions with constraints $|X_2| = 1$ and $|X_2| \leq 1$
enlarges.
When $\frac{P}{\sigma^2} = -5$ dB, the characterized capacity regions with constraints $|X_2| = 1$ and $|X_2| \leq 1$ are the same.
The main reason is that when the received SNR is low, the capacity-achieving distributions on boundary ${\mathbf{B}}$-${\mathbf{C}}$ are 
geometrically characterized by a unit circle for $X_2$.

\begin{figure}[t]
\centering
\includegraphics[width=.9\columnwidth] {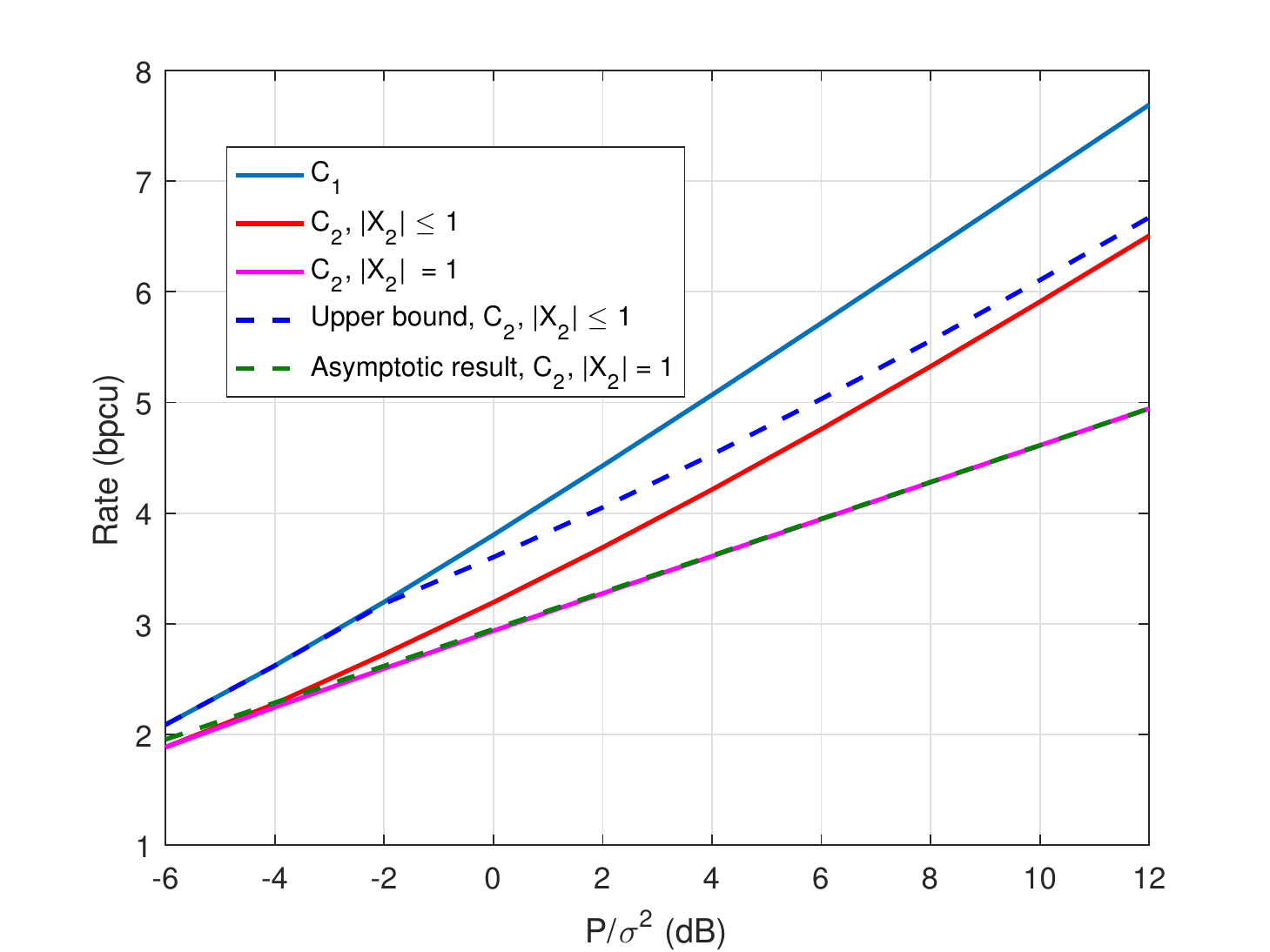}
\caption{Achievable rates versus transmit SNR, i.e., $P/\sigma^2$.}
\label{fig:rate}
\vspace{-1.5em}
\end{figure}

Fig.~\ref{fig:rate} demonstrates the achievable rates for both primary and secondary transmissions versus $P/\sigma^2$.
It can be found that the maximum achievable rate of the primary transmission is always greater than that of the secondary transmission, which is 
consistent with the analysis in Section~\ref{sec:R2}.
Meanwhile, when $|X_2| \leq 1$, the upper bound on the capacity of the secondary transmission shown in~\eqref{eq:upper} becomes
tight when $\frac{P}{\sigma^2}$ is high. The slope of the upper bound is consistent with that of the capacity of the secondary transmission
when $\frac{P}{\sigma^2}$ is high.
In addition, the asymptotic results in~\eqref{eq:asy1} can fit well with the capacity of the secondary transmission given constraint $|X_2| = 1$.
It is obvious that the slope of $C_2$ with $|X_2| \leq 1$ is two times that of $C_2$ with $|X_2| = 1$, which coincides with the
analysis in Section~\ref{sec:Rate2}.

\section {Conclusions}

This paper has focused on the capacity region characterization for RIS-assisted SR when the direct link is blocked. 
The capacity-achieving distributions of the transmitted signals at PTx and RIS have been analyzed. 
Also, the maximum achievable rates for both primary and secondary transmissions as well as their maximum sum rate have been derived.
Furthermore, capacity regions of RIS-assisted SR with two types of passive reflection constraints have been characterized.
Finally, numerical results have been presented to evaluate the performance of RIS-assisted SR.

\appendices

\section{} \label{proof:capacity}

Based on the definition of mutual information, we have $I(X_2;Y|X_1 = x_1)=H({Y}|X_1 = x_1)- H(X_2,X_1 = x_1)$.
It is straightforward to have $ H({Y}|X_2,X_1 = x_1) = H(Z) = \log(\pi e\sigma^2)$.
Then, we focus on the calculation of $H({Y}|X_1 = x_1)$. The polar coordinates of ${Y}$ is given by $(\bar{r},\bar{\psi})$, where $\bar{r}$ and $\bar{\psi}$ are the amplitude and the phase of ${Y}$, respectively. According to Jacobian of transformation in terms of PDF, we have $f_{{y}}(\bar{r},\bar{\psi}|x_1) = \bar{r}p_{{y}}(\mathrm{Re}({y}),\mathrm{Im}({y})|x_1)$, where $p_{{y}}$ is the PDF of ${Y}$ in the Cartesian coordinates. Hence, we have
\begin{align*}
&H({Y} |X_1 = x_1) \\
=& \int_{\mathbb C}p_{{y}}(\mathrm{Re}({y}),\mathrm{Im}({y})|x_1)\log(p_{{y}}(\mathrm{Re}({y}),\mathrm{Im}({y})|x_1))
d{y}\\
\overset{(a)}{=}& -\int_{0}^{+\infty}\int_{-\pi}^{\pi}\bar{r}p_{{y}}(\bar{r}\cos\bar{\psi}, \bar{r}\sin\bar{\psi}|x_1)\\
&~~~~~~~~\qquad\qquad\times\log(
p_{{y}}(\bar{r}\cos\bar{\psi}, \bar{r}\sin\bar{\psi}|x_1)d\bar{\psi} d\bar{r}\\
=& -\int_{0}^{+\infty}\int_{-\pi}^{\pi}f_{{y}}(\bar{r},\bar{\psi}|x_1) \log(\frac{f_{{y}}(\bar{r},\bar{\psi}|x_1)}{\bar{r}})d\bar{\psi} d\bar{r},
\end{align*}
where $(a)$ follows from the Jacobian of transformation in terms of integration. The PDF of ${Y}$ given $X_1$ in terms of the polar coordinates, i.e., $f_{{y}}(\bar{r},\bar{\psi}|x_1)$, is given by
\begin{align}\label{eq:f}
&f_{{y}}(\bar{r},\bar{\psi}|x_1)= \bar{r}p_{y}(\bar{r}\cos\bar{\psi}, \bar{r}\sin\bar{\psi}|x_1)\nonumber \\
=& \bar{r}\int_{x_2}f_2(X_2)p_{y}(\bar{r}\cos\bar{\psi}, \bar{r}\sin\bar{\psi}|x_1,x_2)dx_2\nonumber \\
=&\int_{x_2}f_2(X_2)\frac{\bar{r}}{\pi\sigma^2}\exp\left(-\frac{|\bar{r}e^{j\bar{\psi}}-hx_1x_2|^2}{\sigma^2}\right)dx_2\nonumber \\
=& \int_{-\pi}^{\pi}\frac{\bar{r}}{2\pi^2\sigma^2}\exp\left(-\frac{\bar{r}^2+|h x_1|^2}{\sigma^2}\right)\nonumber \\
&~~~~\times\exp\left(\frac{2\bar{r}|hx_1|\cos(\theta_2+\vartheta-\bar{\psi})}{\sigma^2}\right)d\theta_2 \nonumber \\
=& \frac{\bar{r}}{\pi\sigma^2}\exp\!\left(\!-\frac{\bar{r}^2+|hx_1|^2}{\sigma^2}\!\right)\!I_0\!\left(\!\frac{2\bar{r}|hx_1|}{\sigma^2}\!\right)  
= \frac{\bar{r}\kappa(\bar{r}) }{\pi\sigma^2},
\end{align}
where $\vartheta$ is the phases of $h$, $I_0(x) = \frac{1}{2\pi}\int_{-\pi}^{\pi}e^{x\cos\theta}d\theta$, and $\kappa(\bar{r}) \triangleq \exp\left(-\frac{\bar{r}^2+|hx_1|^2}{\sigma^2}\right)I_0\left(\frac{2\bar{r}|hx_1|}{\sigma^2}\right)$.
 From~\eqref{eq:f}, we can find that the PDF of $f_{{y}}(\bar{r},\bar{\psi}|x_1)$ is irrelevant to the phase of $X_1$ and $h$.
Then, the marginal density $f_{{y}}(\bar{r}|x_1)$ can be obtained by integrating $\bar{\psi}$ out of \eqref{eq:f}:
\begin{align}\label{eq:fff}
f_{{y}}(\bar{r}|x_1)\! &=\! \int_{-\pi}^{\pi}f_{{y}}(\bar{r},\bar{\psi}|x_1)d\bar{\psi} = \frac{2\bar{r}\kappa(\bar{r})}{\sigma^2}.
\end{align}
From \eqref{eq:f} and \eqref{eq:fff}, it is easy to find that $f_{{y}}(\bar{r},\bar{\psi}|x_1) = \frac{1}{2\pi}f_{{y}}(\bar{r}|x_1)$, which means that the amplitude and the phase of ${Y}$ given $X_1$ are independent.
Substituting \eqref{eq:f} and \eqref{eq:fff} into $H({Y} |X_1= x_1)$, we have
\begin{align*}
H({Y}|X_1 = x_1) &= \!-\int_{0}^{+\infty}f_{{y}}(\bar{r}|x_1)\log{\frac{f_{{y}}(\bar{r}|x_1)}{\bar{r}}} d\bar{r}+\log(2\pi) \nonumber \\
&= -\int_{0}^{+\infty}\frac{2\bar{r}\kappa(\bar{r})}{\sigma^2} \log\left(\kappa(\bar{r})\right) d\bar{r}+\log(\pi\sigma^2).
\end{align*}
Thus, the mutual information $I(X_2;Y|X_1 = x_1)$ is given by
\begin{align*}
I(X_2;Y|X_1 = x_1) &=  H({Y}|X_1)- \log(\pi e\sigma^2)\nonumber \\
&= -\int_{0}^{+\infty}\frac{2\bar{r}\kappa(\bar{r})}{\sigma^2} \log\left(\kappa(\bar{r})\right) d\bar{r}-\log(e).
\end{align*}
Therefore, Theorem \ref{theorem:1} can be proved.

\section{}\label{proof:2}
For scheme I, when decoding $X_1$ first, we have $I(X_1;{Y}) = H({Y}) - H({Y}|X_1)$. Due to $X_1X_2\sim \mathcal{CN}(0,P)$, we have $H({Y}) = \log(\pi e(P\tilde h^2+\sigma^2))$.
Next, we focus on the calculation of $H({Y}|X_1)$, which is given by
\begin{align*}
H({Y}|X_1) &\!=\! -\mathbb{E}_{x_1}\!\int_{0}^{+\infty}\!\!\!\int_{-\pi}^{\pi}
f_{{y}}(\bar{r},\bar{\psi}|x_1) \log(\frac{f_{{y}}(\bar{r},\bar{\psi}|x_1)}{\bar r})d\bar{\psi} d\bar{r},
\end{align*}
where $\kappa_{1,1}(\bar{r},\bar{\psi},x_1) \triangleq f_{{y}}(\bar{r},\bar{\psi}|x_1)$ can be written as
\begin{align*}
&\kappa_{1,1}(\bar{r},\bar{\psi},x_1) = \bar{r}p_{y}(\bar{r}\cos\bar{\psi}, \bar{r}\sin\bar{\psi}|x_1)\nonumber \\
=&\int_{x_2}f_2(X_2)\frac{\bar{r}}{\pi\sigma^2}\exp\left(-\frac{|\bar{r}e^{j\bar{\psi}}-\tilde hx_2x_1|^2}{\sigma^2}\right)dx_2\nonumber \\
=& \sum_{n=0}^{\frac{\pi}{\alpha}-1}\frac{\bar{r}\alpha}{\pi^2\sigma^2}\exp\left(-\frac{\bar{r}^2+\tilde h^2|x_1|^2}{\sigma^2}\right)\nonumber \\
&~~~~\times\exp\left(\frac{2\bar{r}\tilde h|x_1|\cos(\theta_1+2n\alpha-\bar{\psi})}{\sigma^2}\right).
\end{align*}
Then, the mutual information of $I(X_1;{Y})$ is given by
\begin{align*}
I(X_1;{Y}) &\!=\! \log(\pi e(P\tilde h^2+\sigma^2))+\mathbb{E}_{x_1}\int_{0}^{+\infty}\!\!\int_{-\pi}^{\pi}\kappa_{1,1}(\bar{r},\bar{\psi},x_1)\\
&\qquad\qquad\times\log(\frac{\kappa_{1,1}(\bar{r},\bar{\psi},x_1)}{\bar r})d\bar{\psi}  d\bar{r}.
\end{align*}
Since $I(X_2;{Y}|X_1) = H({Y}|X_1)-H({Y}|X_1,X_2)$ and $H({Y}|X_1,X_2) = \log(\pi e\sigma^2)$, we have
\begin{align*}
I(X_2;{Y}|X_1)  &= - \mathbb{E}_{x_1}\int_{0}^{+\infty}\!\!\int_{-\pi}^{\pi}\kappa_{1,1}(\bar{r},\bar{\psi},x_1)\\
&\qquad\quad\times\log(\frac{\kappa_{1,1}(\bar{r},\bar{\psi},x_1)}{\bar r})d\bar{\psi}  d\bar{r}-\log(\pi e\sigma^2).
\end{align*}

The calculations of mutual information for scheme II and the other decoding order are similar to the above calculations and thus are omitted here.

\section{}\label{proof:feature}
Due to the multiplicative nature between $X_1$ and $X_2$, the effect of unknown $X_2$ on mutual information 
is the same as that of the unknown channel state information. Thus, the structures of the existence proof below and Appendix~\ref{proof:finite}
follow~\cite{abou2001capacity}, 
but the details are different since the objective function in problem $\textbf{P2}$ is more difficult to deal with.
Considering the average power constraint of $X_1$, according to Lagrangian Theorem, problem $\textbf {P2}$ can be recast as
\begin{align}\label{eq:ne12}
\mathop {\min}\limits_{\lambda}\mathop {\max }\limits_{f_a}~~&\mu_1 H(Y)+(\mu_2-\mu_1)H(Y|A) \nonumber \\
&-\lambda\left(\int_{0}^{\infty}f_aa^2d a-P\right),
\end{align}
where $\lambda\geq 0$ is a Laplace multiplier. 
Let $T$ be a functional on a convex set $\mathcal F_a$, where $T(f_a^{*}) = \mu_1H(Y;f_a^{*})+(\mu_2-\mu_1)H(Y|A;f_a^{*})-\lambda\int_{0}^{\infty}f_a^*a^2d a$ with $f_a^{*} \in \mathcal F_a$. Then, the weak derivative of $T(f_a^{*})$ is given by
\begin{align*}
T_{f_a^{*}}^{'}(f_a) = & \lim_{\theta\rightarrow 0}\frac{T((1-\theta)f_a^{*}+\theta f_a)-T(f_a^{*})}{\theta},
\end{align*}
where $\theta \in [0,1]$ and $f_a$ is an arbitrary PDF in $\mathcal F_a$. Define $f_a^{\theta} = (1-\theta)f_a^{*}+\theta f_a$, and then, we have
\begin{align*}
&T(f_a^{\theta})-T(f_a^{*}) \\
=& \mu_1\int_0^{+\infty} f_a^{\theta}\omega(a;f_a^{\theta})da-\mu_1\int_0^{+\infty} f_a^{*}\omega(a;f_a^{*})da\\
&~~+\int_0^{+\infty} (f_a^{\theta}-f_a^{*})((\mu_2-\mu_1)H(Y|A=a)-\lambda a^2)da\\
=&\mu_1\int_0^{+\infty} f_a^*(\omega(a;f_a^{\theta})-\omega(a;f_a^*))da+\theta\left(\int_0^{+\infty} (f_a-f_a^{*})\right.\\
&\quad\times(\mu_1\omega(a;f_a^{\theta})+(\mu_2-\mu_1)H(Y|A=a)-\lambda a^2)da).
\end{align*}
Accordingly
\begin{align*}
T_{f_a^{*}}^{'}(f_a)& = \int_0^{+\infty} (f_a-f_a^{*})(\mu_1\omega(a;f_a^{*})\\
&~~~~+(\mu_2-\mu_1)H(Y|A=a)-\lambda a^2)da.
\end{align*}
According to~\cite{luenberger1997optimization}, if $T$ achieves its maximum at $f_a^{*}$, then we have 
$T_{f_a^{*}}^{'}(f_a)\leq 0$ for all $f_a \in \mathcal F_a$, yielding
\begin{align}\label{eq:cond1}
&\int_0^{+\infty}\!\! f_a\!\left(\omega(a;f_a^{*})+\!\left(\!\frac{\mu_2}{\mu_1}-1\!\right)\!H(Y|A=a)-\frac{\lambda}{\mu_1} a^2\!\right)da \nonumber \\
\leq&\!\int_0^{+\infty}\!\!f_a^{*}\!\left(\omega(a;f_a^{*})+\!\left(\!\frac{\mu_2}{\mu_1}-1\!\right)\!H(Y|A=a)\!-\!\frac{\lambda}{\mu_1} a^2\!\right)da.
\end{align}
The right-hand side (RHS) of~\eqref{eq:cond1} can be represented by $T_0 = \int_0^{\infty} f_a^*\omega(a;f_a^{*})da+(\frac{\mu_2}{\mu_1}-1)\int_0^{\infty} f_a^*H(Y|A=a)da-\frac{\lambda}{\mu_1} P$. Note that if $\int_0^{+\infty}f_a^{*}a^2 da$ is less than $P$, $\lambda$ is zero according to Karush–Kuhn–Tucker conditions, which indicates that~\eqref{eq:cond1} remains true.

We use contradictions to prove~\eqref{eq:condtion1} in Theorem~\ref{theorem:feature}. If \eqref{eq:condtion1} is false, there exist an $\tilde {a}$ such that
\begin{align*}
\omega(\tilde a;f_a^{*}) > T_0-\left(\frac{\mu_2}{\mu_1}-1\right)H(Y|A=\tilde a)+\frac{\lambda}{\mu_1}\tilde a^2.
\end{align*} 
Since~\eqref{eq:cond1} holds for all $f_a \in \mathcal F_a$, we assume $f_a = \delta (A-\tilde a)$, and then we have
\begin{align*}
\int_0^{+\infty}\!\!\! f_a\!\left(\!\omega(a;f_a^{*})+\!\left(\!\frac{\mu_2}{\mu_1}-1\right)H(Y|A={a})\!-\!\frac{\lambda}{\mu_1} {a}^2\!\right)da 
> T_0,
\end{align*}
which is contradicts~\eqref{eq:cond1}. Thus, \eqref{eq:condtion1} is valid.

Then, we focus on the proof of~\eqref{eq:condtion2} in Theorem~\ref{theorem:feature}, which can also be proved by contradictions.
Specifically, we assume there exists $\tilde a \in E_0$ such that 
\begin{align*}
\omega(\tilde a;f_a^{*}) < T_0-\left(\frac{\mu_2}{\mu_1}-1\right)H(Y|A=\tilde a)+\frac{\lambda}{\mu_1}\tilde a^2.
\end{align*} 
Then, by using the definition of a point of increase of CDF, the neighborhood $\tilde E$ of $\tilde a$ satisfies 
$\int_{\tilde E} f_a^* da  = \varrho >0$. Hence, we have
\begin{align*}
T_0 &= \int_0^{\infty}\!\! f_a^*\!\left(\!\omega(a;f_a^{*})+\!\left(\!\frac{\mu_2}{\mu_1}-1\!\right)\!H(Y|A=a)\!\right)da\!-\!\frac{\lambda}{\mu_1} P\\
& = \int_{\tilde E} \!f_a^*\left(\omega(a;f_a^{*})+\!\left(\!\frac{\mu_2}{\mu_1}-1\!\right)\!H(Y|A=a)\right)da \\
& +\int_{E_0-\tilde E} \!f_a^*\!\left(\!\omega(a;f_a^{*})+\!\left(\!\frac{\mu_2}{\mu_1}-1\!\right)\!H(Y|A=a)\right)\!da\!-\!\frac{\lambda}{\mu_1} P\\
& < \varrho \left(T_0+\frac{\lambda}{\mu_1} P\right) +(1-\varrho) \left(T_0+\frac{\lambda}{\mu_1} P\right) -\frac{\lambda}{\mu_1} P= T_0,
\end{align*} 
which is a contradiction. Thus, \eqref{eq:condtion2} holds.

\section{}\label{proof:finite}
Assume that there exists a continuous optimal $f_a^*$ such that \eqref{eq:condtion2} holds.
By letting $\varpi \triangleq \frac{\tilde h^2 a^2}{\sigma^2}$ and based on Appendix A, we have
\begin{align}\label{eq:omega}
\omega(A;f_a^{*})=&- \int_0^{+\infty}\frac{2\bar{r}}{\sigma^2}\exp\!\left(\!-\frac{\bar{r}^2}{\sigma^2}-\varpi\right)\nonumber\\
&\times I_0\left(\frac{2\bar{r}\sqrt{\varpi}}{\sigma}\!\right)\log f_y(\bar{r};f_a^*)d\bar{r} +\log (2\pi).
\end{align}
By using Laplace transform with respect to $\varpi$ over~\eqref{eq:omega}, we have
\begin{align*}
&\int_0^{+\infty} e^{-\varpi s} (\omega(A;f_a^{*})-\log(2\pi)) d \varpi\\
=& -\int_0^{+\infty}\int_0^{+\infty}\frac{2\bar{r}}{\sigma^2}\exp\!\left(\!-\frac{\bar{r}^2}{\sigma^2}-\varpi-\varpi s\right)\\
&~\qquad \times I_0\left(\frac{2\sqrt{\varpi}\bar{r}}{\sigma}\!\right)\log f_y(\bar{r};f_a^*)d\bar{r}d \varpi.
\end{align*}
By using~\cite[pp. 697]{gradshteyn2014table}
\begin{align*}
\int_0^{+\infty} \exp(-\alpha x)I_0(2\sqrt{\beta x})d x = \frac{1}{\alpha}\exp\left(\frac{\beta}{\alpha}\right),
\end{align*}
we have
\begin{align*}
&\int_0^{+\infty} e^{-\varpi s} (\omega(A;f_a^{*})-\log(2\pi)) d \varpi\\
= & -\int_0^{+\infty}\frac{2\bar{r}}{\sigma^2\left(
1+s\right)}\exp\!\left(-\frac{\bar{r}^2}{\sigma^2}\left(\frac{s}{1+s}\right)\right)\log f_y(\bar{r};f_a^*)d\bar{r}\\
= & -\frac{1}{\sigma^2\left(
1+s\right)}\int_0^{+\infty}\exp\!\left(-\frac{\xi}{\sigma^2}\left(\frac{s}{1+s}\right)\right)\log f_\xi(\xi;f_a^*)d\xi,
\end{align*}
where $\xi\triangleq \bar{r}^2 $.

By using Laplace transform to the RHS of~\eqref{eq:condtion2} with respect to $\varpi$, we have
\begin{align*}
&\int_0^{+\infty} e^{-\varpi s} \left(T_0-\left(\frac{\mu_2}{\mu_1}-1\right)H(Y|\varpi)+\frac{\lambda\sigma^2\varpi}{\mu_1\tilde h^2}\right) d \varpi\\
=& \frac{T_0}{s}+\frac{\lambda\sigma^2}{\mu_1 \tilde h^2s^2}-\left(\frac{\mu_2}{\mu_1}-1\right)\mathcal H(s),
\end{align*}
where $\mathcal H(s)\triangleq\mathcal{L}(H(Y|\varpi))$. By letting $\beta = \frac{s}{\sigma^2\left(
1+s\right)}$, the results of using Laplace transform and multiplied by $s$ on both sides of~\eqref{eq:condtion2} becomes
\begin{align}\label{eq:beta}
& -\beta\int_0^{+\infty}\exp\!\left(-\xi\beta \right)\log f_\xi(\xi;f_1^*)d\xi\nonumber\\
= &~ T_1+\!\frac{\lambda(1-\beta\sigma^2)}{\mu_1\tilde h^2\beta}\!-\!\left(\!\frac{\mu_2}{\mu_1}\!-\!1\!\right)\!
\frac{\beta\sigma^2}{1\!-\!\beta\sigma^2}\mathcal H\!\left(\!\frac{\beta\sigma^2}{1\!-\!\beta\sigma^2}\!\right),
\end{align}
where $T_1 = T_0-\log(2\pi)$
Dividing~\eqref{eq:beta} by $\beta$ yields
\begin{align}\label{eq:inverse}
&\int_0^{+\infty}\exp\left(-\xi\beta \right)\log f_\xi(\xi;f_1^*)d\xi\nonumber\\
= &-\frac{T_1}{\beta}-\frac{\lambda(1\!-\!\beta\sigma^2)}{\mu_1\tilde h^2\beta^2}+\!\left(\!\frac{\mu_2}{\mu_1}-1\!\right)\!\frac{\sigma^2}{1-\beta\sigma^2}\mathcal H\!\left(\!\frac{\beta\sigma^2}{1-\beta\sigma^2}\right).
\end{align}
It is obvious that the left-hand side (LHS) of~\eqref{eq:inverse} is the unilateral Laplace transform of function $\log f(\xi;f_1^*)$, while the RHS of~\eqref{eq:inverse} can be recognized as the Laplace transform of
\begin{align*}
&\! -\!T_1+\!\frac{\lambda\sigma^2}{\mu_1\tilde h^2}\!-\!\frac{\lambda \xi}{\mu_1\tilde h^2}+\left(\!\frac{\mu_2}{\mu_1}\!-\!1\!\right)\!\!\mathcal{L}^{-1}\!\!\left(\!\frac{\sigma^2}{1-\beta\sigma^2}\mathcal H\!\left(\!\frac{\beta\sigma^2}{1-\beta\sigma^2}\!\right)\!;\xi\!\right)\!,
\end{align*}
which implies that
\begin{align}\label{eq:pdfy}
& f_\xi(\xi;f_a^*) = \exp\left(-T_1+\frac{\lambda\sigma^2}{\mu_1\tilde h^2}-\frac{\lambda \xi}{\mu_1\tilde h^2}\right.\nonumber\\
&\quad\left.+\left(\frac{\mu_2}{\mu_1}-1\right)\mathcal{L}^{-1}\left(\frac{\sigma^2}{1-\beta\sigma^2}\mathcal H\left(\frac{\beta\sigma^2}{1-\beta\sigma^2}\right)\right);\xi\right).
\end{align}
From~\eqref{eq:pdfy}, one can find if $\mu_1 = \mu_2$, we have $f(\xi;f_a^*) = \exp\left(-T_1+\frac{\lambda\sigma^2}{\mu_1\tilde h^2}-\frac{\lambda \xi}{\mu_1\tilde h^2}\right)$ such that the received signal $Y$ follows a Gaussian distribution, which is achieved when $X_1X_2$ follows a Gaussian distribution. 
However, the points on boundary ${\mathbf{B}}$-${\mathbf{C}}$ require $\mu_2>\mu_1$. Thus, we need to consider the effect of $\mathcal{L}^{-1}\left(\frac{\sigma^2}{1-\beta\sigma^2}\mathcal H(\frac{\beta\sigma^2}{1-\beta\sigma^2});\xi\right)$.

First, we focus on $\mathcal H(s)$. Based on~\eqref{eq:asy1}, we have
\begin{align}\label{eq:Lap}
&\mathcal H(s) =\mathcal{L}(H(Y|\varpi))= \mathcal{L}\left(\!\varpi(u(\varpi)-u(\varpi-1))\right) \nonumber\\
&\qquad\quad+\mathcal{L}\left(\frac{1}{2}\log\left(\frac{4\pi\varpi}{e}\right) u(\varpi-1)+\log(\pi e\sigma^2)\right)\nonumber\\
= &\!\frac{1}{s^2}\!-\! \frac{e^{-s}}{s^2}\!-\!\frac{e^{-s}}{s} \!-\!\frac{1}{2s}\mathrm{Ei}(-s)\!+\!\log\!\left(\!\frac{4\pi}{e}\!\right)\!\frac{e^{-s}}{2s}\!+\!\frac{\log(\pi e\sigma^2)}{s},
\end{align}
where $u(\varpi)$ is a unit step function. 
Then, one of terms of $\mathcal{L}^{-1}\left(\frac{\sigma^2}{1-\beta\sigma^2}\mathcal H(\frac{\beta\sigma^2}{1-\beta\sigma^2});\xi\right)$
corresponds to the first term in~\eqref{eq:Lap}, which is given by
\begin{align*}
&\mathcal{L}^{-1}\left(\frac{1-\beta\sigma^2}{\beta^2\sigma^2};\xi\right)  = \frac{\xi}{\sigma^2} - 1.
\end{align*}

Due to $\frac{\mu_2}{\mu_1}-1>0$, the PDF in~\eqref{eq:pdfy} does not converge. Hence $f_{\xi}(\xi;f_1^*)$ in~\eqref{eq:pdfy} cannot be a probability density. Thus, the increase points set $E_0$ of the optimal CDF $F_a^*$ is finite, which implies that the optimal PDF $f_a^*$ is discrete.


\section{}\label{proof:1}

The distribution of $X_1$ is $f_1(X_1=ae^{j\theta_1}) = \sum_{m = 1}^{M}p_m \delta(a-a_m)\delta(\theta_1)$ and the distribution of $X_2$ is $f_2(X_2 = e^{j\theta_2}) = \frac{1}{2\pi}$. Based on Appendix~\ref{proof:capacity}, we have
\begin{align*}
&H({Y}|X_1)= \int_{x_1} f_1(x_1)H({Y} |X_1 = x_1) dx_1 \nonumber \\
=& \!-\!\sum_{m= 1}^{M}\!p_m\!\int_{0}^{+\infty}\!\!\frac{2\bar{r}\bar{\kappa}(\bar{r},\!a_m)}{\sigma^2} \!\log\!\left(\!\frac{2\bar{\kappa}(\bar{r},\!a_m)}{\sigma^2}\!\right)\! d\bar{r}\!+\!\log(2\pi),
\end{align*}
where $\bar{\kappa}(\bar{r},a_m) =\exp\left(-\frac{\bar{r}^2+\tilde h^2a_m^2}{\sigma^2}\!\right)I_0\left(\frac{2\bar{r}\tilde ha_m}
{\sigma^2}\right)$. Then, $I(X_2;Y|X_1)$ is calculated by the definition of mutual information.

Next, we focus on the calculation of $H({Y})$, which is given by $H({Y}) = -\int_{0}^{+\infty}$
$\int_{-\pi}^{\pi}f_{{y}}(\bar{r},\bar{\psi})\log(\frac{f_y(\bar{r},\bar{\psi})}{\bar{r}})d\bar{\psi} d\bar{r}$ and
\begin{align*}
f_y(\bar{r},\bar{\psi})\!&=\! \bar{r}\!\int_{x_1}\!\int_{x_2}f_1(X_1)f_2(X_2)\\
&~~~\qquad\times\frac{1}{\pi\sigma^2}\exp\left(-\frac{|\bar{r}e^{j\psi}-\tilde hx_1x_2|^2}{\sigma^2}\right)dx_1dx_2 \nonumber\\
&= \sum_{m = 1}^{M}\frac{p_m \bar{r}}{2\pi^2\sigma^2}\int_{-\pi}^{\pi}\exp\!\left(-\frac{|\bar{r}e^{j\psi}-\tilde ha_m e^{j\theta_2}|^2}{\sigma^2}\!\right)d\theta_2\nonumber \\
&=\sum_{m = 1}^{M}\frac{p_m \bar{r}}{2\pi^2\sigma^2}\exp\left(-\frac{\bar{r}^2+\tilde h^2a_m^2}{\sigma^2}\right)\int_{-\pi}^{\pi}\nonumber \\
&~~\quad\qquad\times\exp\left(\frac{2\bar{r}\tilde ha_m\cos(\theta_2-\bar{\psi})}{\sigma^2}\!\right)d\theta_2\nonumber\\
&= \sum_{m = 1}^{M}\frac{p_m \bar{r}}{\pi\sigma^2}\exp\left(-\frac{\bar{r}^2+\tilde h^2a_m^2}{\sigma^2}\right)I_0\left(\frac{2\bar{r}\tilde ha_m}{\sigma^2}\right).
\end{align*}
Since $f_{{y}}(\bar{r})$ satisfies $f_{{y}}(\bar{r},\bar{\psi}) = \frac{1}{2\pi}f_{{y}}(\bar{r})$, we have
$H({Y})= -\int_{0}^{+\infty}f_{{y}}(\bar{r})\log{\frac{f_{{y}}(\bar{r})}{\bar{r}}} d\bar{r}+\log(2\pi)$.
By the definition of mutual information, we can have $I(X_1;Y)$ and $I(X_2;Y|X_1)$.

\section{}\label{proof:finite2}

Following the derivations of Appendix~\ref{proof:feature}, one can find that with constraint $|X_2|\leq 1$, \eqref{eq:condtion1} and~\eqref{eq:condtion2} still hold.
Similar to Appendix~\ref{proof:finite}, we assume that there exists a continuous optimal $f_a^*$ such that \eqref{eq:condtion2} holds.
Then, for the case $|X_2|\leq 1$, the marginal entropy $\omega(A;f_a^*)$ can be rewritten as
\begin{align}\label{eq:omega2}
\omega(A;f_a^{*})=&- \int_{0}^{+\infty }f_r(r)\int_0^{+\infty}\frac{2\bar{r}}{\sigma^2}\exp\!\left(\!-\frac{\bar{r}^2}{\sigma^2}-r^2\varpi\right)\nonumber\\
&\times I_0\left(\frac{2\bar{r}r\sqrt{\varpi}}{\sigma}\!\right)\log f_y(\bar{r};f_a^*)d\bar{r}dr+\log(2\pi).
\end{align}
By using Laplace transform with respect to $\varpi$ over~\eqref{eq:omega2}, we have
\begin{align}\label{eq:cond2}
&\int_0^{+\infty} e^{-\varpi s}(\omega(A;f_a^{*})-\log(2\pi)) d \varpi\nonumber\\
= & -\!\!\! \int_{0}^{+\infty }\!\!\!\!\!\frac{f_r(r)}{\sigma^2\!\left(
r^2+s\right)}\!\int_0^{+\infty}\!\!\!\exp\!\left(\!-\frac{\xi}{\sigma^2}\!\left(\!\frac{s}{r^2+s}\!\right)\!\right)
\!\log f_{\xi}(\xi;f_a^*)d\xi dr \nonumber\\
= & - \int_{0}^{+\infty }\frac{f_r(r)}{\sigma^2\left(
r^2+s\right)}F\left(\frac{s}{(r^2+s)\sigma^2}\right)dr\nonumber\\
=& - \int_{0}^{+\infty }\frac{f_r(r)\beta(r,s)}{s} F\left(\beta(r,s)\right)dr\nonumber\\
\overset{(a)}{=} &~  \frac{T_0-\log(2\pi)}{s}+\frac{\lambda\sigma^2}{\mu_1 \tilde h^2s^2}-\left(\frac{\mu_2}{\mu_1}-1\right)\mathcal H(s),
\end{align}
where $ F(s) = \mathcal{L}(\log f_\xi(\xi;f_a^*))$, $\beta(r,s) \triangleq \frac{s}{(r^2+s)\sigma^2}$, and $(a)$ holds by using Laplace transform to the RHS of~\eqref{eq:condtion2}.
If there exists $f_a^*$ such that~\eqref{eq:cond2} holds, there will exist $F_1\left(s\right)$, $F_2\left(s\right)$, and $F_3\left(s\right)$ such that $\int_{0}^{+\infty }f_r(r)\beta(r,s) F_1\left(\beta(r,s)\right)dr=-T_0+\log(2\pi)$, $\int_{0}^{+\infty }f_r(r)\beta(r,s) F_2\left(\beta(r,s)\right)dr=-\frac{\lambda\sigma^2}{\mu_1 \tilde h^2s}$, and $\int_{0}^{+\infty }f_r(r)\beta(r,s) F_3\left(\beta(r,s)\right)dr=\left(\frac{\mu_2}{\mu_1}-1\right)s\mathcal H(s)$, where $F\left(\beta(r,s)\right) = F_1\left(\beta(r,s)\right)+F_2\left(\beta(r,s)\right)+F_3\left(\beta(r,s)\right)$.

It is straightforward to obtain $F_1\left(s\right) = \frac{-T_0+\log(2\pi)}{s}$. 
Assume that there exists a unique solution $F_2^*\left(s\right)$ such that 
$\int_{0}^{+\infty }f_r(r)\beta(r,s) F_2\left(\beta(r,s)\right)dr=-\frac{\lambda\sigma^2}{\mu_1 \tilde h^2s}$. 
To enable this equation, we have $F_2^*\left(\beta(r,s)\right) = \frac{-\lambda \sigma^4 (r^2+s)}{\mu_1\tilde h^2 s^2}$, which yields $F_2^*\left(s\right) = \frac{-\lambda (1-s\sigma^2)}{\mu_1 \tilde h ^2s^2r^2}$. 
If the amplitude of $X_2$ is constant, i.e., $|X_2| = r_0$, we have $F_2^*\left(s\right) = \frac{-\lambda (1-s\sigma^2)}{\mu_1 \tilde h ^2s^2r_0^2}$, 
which is a unique solution and consistent with the results in Appendix~\ref{proof:finite}.
In this case, the optimal $f_a^*$ is discrete.
If the amplitude of $X_2$ is not constant, we can find that the derived $F_2^*(s)$ is related to $r$, which is not a unique solution. Thus, there is no 
solution for~\eqref{eq:cond2}.
This indicates that the amplitude of $X_2$ requires to be constant to enable $X_1X_2$ follows a Gaussian distribution, which coincides with 
the results in Section~\ref{sec:op}. 

In summary, there is no continuous optimal PDF $f_a^*$ to enable~\eqref{eq:condtion2} to hold. Hence, under constraint $|X_1|\leq 1$,
the optimal PDF $f_a^*$ is still discrete.
For a discrete $f_a^*$, it is easy to have that $f_r^*$ is also discrete by following the proof structures of Appendix~\ref{proof:feature}
and Appendix~\ref{proof:finite}.


\vspace{-0.5em}
\begin{thebibliography}{10}
\providecommand{\url}[1]{#1}
\csname url@samestyle\endcsname
\providecommand{\newblock}{\relax}
\providecommand{\bibinfo}[2]{#2}
\providecommand{\BIBentrySTDinterwordspacing}{\spaceskip=0pt\relax}
\providecommand{\BIBentryALTinterwordstretchfactor}{4}
\providecommand{\BIBentryALTinterwordspacing}{\spaceskip=\fontdimen2\font plus
\BIBentryALTinterwordstretchfactor\fontdimen3\font minus
  \fontdimen4\font\relax}
\providecommand{\BIBforeignlanguage}[2]{{%
\expandafter\ifx\csname l@#1\endcsname\relax
\typeout{** WARNING: IEEEtran.bst: No hyphenation pattern has been}%
\typeout{** loaded for the language `#1'. Using the pattern for}%
\typeout{** the default language instead.}%
\else
\language=\csname l@#1\endcsname
\fi
#2}}
\providecommand{\BIBdecl}{\relax}
\BIBdecl

\bibitem{zhang2023ICC}
Q.~Zhang, H.~Zhou, Y.-C. Liang, W.~Zhang, and H.~V. Poor, ``On the capacity
  region of reconfigurable intelligent surface assisted symbiotic radios,'' in
  \emph{Proc. {IEEE} Int. Conf. Commun. (ICC)}.\hskip 1em plus 0.5em minus
  0.4em\relax IEEE, 2023, pp. 1--6.

\bibitem{matti2019key}
M.~Latva-aho and K.~Leppanen, ``Key drivers and research challenges for 6{G}
  ubiquitous wireless intelligence,'' University of Oulu, While Paper, 2019.
  Available: http://urn.fi/urn:isbn:9789526223544.

\bibitem{giordani2020toward}
M.~Giordani, M.~Polese, M.~Mezzavilla, S.~Rangan, and M.~Zorzi, ``Toward {6G}
  networks: Use cases and technologies,'' \emph{IEEE Commun. Mag.}, vol.~58,
  no.~3, pp. 55--61, 2020.

\bibitem{xiaohutowards}
X.~You \emph{et~al.}, ``Towards {6G} wireless communication networks: Vision,
  enabling technologies, and new paradigm shifts,'' \emph{Sci. China Inf.
  Sci.}, vol.~64, no.~1, pp. 1--74, 2021.

\bibitem{ge2023deep}
J.~Ge, Y.-C. Liang, L.~Zhang, R.~Long, and S.~Sun, ``Deep reinforcement
  learning for distributed dynamic coordinated beamforming in massive mimo
  cellular networks,'' \emph{IEEE Trans. Wireless Commun.}, 2023, DOI:
  10.1109/TWC.2023.3314930.

\bibitem{liang2020symbiotic}
Y.-C. Liang, Q.~Zhang, E.~G. Larsson, and G.~Y. Li, ``Symbiotic radio:
  Cognitive backscattering communications for future wireless networks,''
  \emph{IEEE Trans. Cogn. Commun. Netw.}, vol.~6, no.~4, pp. 1242--1255, 2020.

\bibitem{liu2013ambient}
V.~Liu, A.~Parks, V.~Talla, S.~Gollakota, D.~Wetherall, and J.~R. Smith,
  ``Ambient backscatter: Wireless communication out of thin air,'' in
  \emph{Proc. of ACM SIGCOMM}, vol.~43, no.~4.\hskip 1em plus 0.5em minus
  0.4em\relax Hong Kong, China: ACM, Aug. 2013, pp. 39--50.

\bibitem{Boyer2014Backscatter}
C.~Boyer and S.~Roy, ``Backscatter communication and {RFID}: {Coding}, energy,
  and {MIMO} analysis,'' \emph{IEEE Trans. Commun.}, vol.~62, no.~3, pp.
  770--785, Mar. 2014.

\bibitem{yang2018modulation}
G.~Yang, Y.-C. Liang, R.~Zhang, and Y.~Pei, ``Modulation in the air:
  Backscatter communication over ambient {OFDM} carrier,'' \emph{IEEE Trans.
  Commun.}, vol.~66, no.~3, pp. 1219--1233, Mar. 2018.

\bibitem{zhang2022mutualistic}
Q.~Zhang, Y.-C. Liang, H.-C. Yang, and H.~V. Poor, ``Mutualistic mechanism in
  symbiotic radios: When can the primary and secondary transmissions be
  mutually beneficial?'' \emph{IEEE Trans. Wireless Commun.}, vol.~21, no.~10,
  pp. 8036--8050, 2022.

\bibitem{chen2020vision}
S.~Chen, Y.-C. Liang, S.~Sun, S.~Kang, W.~Cheng, and M.~Peng, ``Vision,
  requirements, and technology trend of {6G}: How to tackle the challenges of
  system coverage, capacity, user data-rate and movement speed,'' \emph{IEEE
  Wireless Commun.}, vol.~27, no.~2, pp. 218--228, 2020.

\bibitem{9178307}
L.~Bariah, L.~Mohjazi, S.~Muhaidat, P.~C. Sofotasios, G.~K. Kurt,
  H.~Yanikomeroglu, and O.~A. Dobre, ``A prospective look: Key enabling
  technologies, applications and open research topics in {6G} networks,''
  \emph{IEEE Access}, vol.~8, pp. 174\,792--174\,820, 2020.

\bibitem{yang2018cooperative}
G.~Yang, Q.~Zhang, and Y.-C. Liang, ``Cooperative ambient backscatter
  communications for green {Internet-of-Things},'' \emph{IEEE Internet Things
  J.}, vol.~5, no.~2, pp. 1116--1130, Apr. 2018.

\bibitem{chen2020stochastic}
X.~Chen, H.~V. Cheng, K.~Shen, A.~Liu, and M.-J. Zhao, ``Stochastic transceiver
  optimization in multi-tags symbiotic radio systems,'' \emph{IEEE Internet
  Things J.}, vol.~7, no.~9, pp. 9144--9157, 2020.

\bibitem{zhang2019constellation}
Q.~Zhang, H.~Guo, Y.-C. Liang, and X.~Yuan, ``Constellation learning-based
  signal detection for ambient backscatter communication systems,'' \emph{IEEE
  J. Sel. Areas Commun.}, vol.~37, no.~2, pp. 452--463, 2019.

\bibitem{guo2019exploiting}
H.~Guo, Q.~Zhang, S.~Xiao, and Y.-C. Liang, ``Exploiting multiple antennas for
  cognitive ambient backscatter communication,'' \emph{IEEE Internet Things
  J.}, vol.~6, no.~1, pp. 765--775, 2019.

\bibitem{long2019symbiotic}
R.~Long, Y.-C. Liang, H.~Guo, G.~Yang, and R.~Zhang, ``Symbiotic radio: A new
  communication paradigm for passive internet-of-things,'' \emph{IEEE Internet
  Things J.}, vol.~7, pp. 1350--1363, 2020.

\bibitem{chu2020resource}
Z.~Chu, W.~Hao, P.~Xiao, M.~Khalily, and R.~Tafazolli, ``Resource allocations
  for symbiotic radio with finite block length backscatter link,'' \emph{IEEE
  Internet Things J.}, vol.~7, no.~9, pp. 8192--8207, 2020.

\bibitem{10499212}
J.~Wang, Y.-C. Liang, and S.~Sun, ``Multi-user multi-iot-device symbiotic
  radio: A novel massive access scheme for cellular iot,'' \emph{IEEE Trans.
  Wireless Commun.}, 2024, DOI: 10.1109/TWC.2024.3385530.

\bibitem{zhang2021reconfigurable}
Q.~Zhang, Y.-C. Liang, and H.~V. Poor, ``Reconfigurable intelligent surface
  assisted {MIMO} symbiotic radio networks,'' \emph{IEEE Trans. Commun.},
  vol.~69, no.~7, pp. 4832--4846, 2021.

\bibitem{ma2020large}
T.~Ma, Y.~Xiao, X.~Lei, P.~Yang, X.~Lei, and O.~A. Dobre, ``Large intelligent
  surface assisted wireless communications with spatial modulation and antenna
  selection,'' \emph{IEEE J. Sel. Areas Commun.}, vol.~38, no.~11, pp.
  2562--2574, 2020.

\bibitem{liang2022backscatter}
Y.-C. Liang, Q.~Zhang, J.~Wang, R.~Long, H.~Zhou, and G.~Yang, ``Backscatter
  communication assisted by reconfigurable intelligent surfaces,'' \emph{Proc.
  IEEE}, vol. 110, no.~9, pp. 1339--1357, 2022.

\bibitem{liu2019symbol}
R.~Liu, H.~Li, M.~Li, and Q.~Liu, ``Symbol-level precoding design for
  intelligent reflecting surface assisted multi-user {MIMO} systems,'' in
  \emph{Proc. {IEEE} Int. Conf. Wireless Commun. Signal Process. (WCSP)}.\hskip
  1em plus 0.5em minus 0.4em\relax IEEE, 2019, pp. 1--6.

\bibitem{zhou2023modulation}
H.~Zhou, B.~Cai, Q.~Zhang, R.~Long, Y.~Pei, and Y.-C. Liang, ``Modulation
  design and optimization for {RIS}-assisted symbiotic radios,'' \emph{arXiv
  preprint arXiv:2311.01167}, 2023.

\bibitem{10336749}
H.~Zhou, Q.~Zhang, Y.-C. Liang, and Y.~Pei, ``Assistance-transmission tradeoff
  for {RIS}-assisted symbiotic radios,'' \emph{IEEE Trans. Wireless Commun.},
  2023, DOI:10.1109/TWC.2023.3335111.

\bibitem{wang2023mutualistic}
Y.~Wang, Q.~Zhang, H.~Zhou, and Y.-C. Liang, ``Mutualistic mechanism for
  ris-assisted symbiotic radios: How many reflecting elements are required?''
  in \emph{Proc. of IEEE Glob. Commun. Conf. (GLOBECOM)}.\hskip 1em plus 0.5em
  minus 0.4em\relax IEEE, 2023, pp. 4424--4429.

\bibitem{karasik2021single}
R.~Karasik, O.~Simeone, M.~Di~Renzo, and S.~Shamai, ``{Single-RF} multi-user
  communication through reconfigurable intelligent surfaces: An
  information-theoretic analysis,'' in \emph{IEEE Int. Symp. Inf. Theory
  (ISIT)}, 2021, pp. 2352--2357.

\bibitem{cheng2021degree}
H.~V. Cheng and W.~Yu, ``Degree-of-freedom of modulating information in the
  phases of reconfigurable intelligent surface,'' \emph{IEEE Trans. Inf.
  Theory}, vol.~70, no.~1, pp. 170--188, 2024.

\bibitem{hua2021novel}
M.~Hua, Q.~Wu, L.~Yang, R.~Schober, and H.~V. Poor, ``A novel wireless
  communication paradigm for intelligent reflecting surface based symbiotic
  radio systems,'' \emph{IEEE Trans. Signal Process.}, vol.~70, pp. 550--565,
  2021.

\bibitem{zhou2022cooperative}
H.~Zhou, X.~Kang, Y.-C. Liang, S.~Sun, and X.~Shen, ``Cooperative beamforming
  for reconfigurable intelligent surface-assisted symbiotic radios,''
  \emph{IEEE Trans. Veh. Technol.}, vol.~71, no.~11, pp. 11\,677--11\,692,
  2022.

\bibitem{zhang2020sr}
Q.~Zhang, Y.-C. Liang, and H.~V. Poor, ``Symbiotic radio: A new application of
  largeintelligent surface/antennas {(LISA)},'' \emph{Proc. IEEE Wireless
  Commun. Netw. Conf. (WCNC)}, May 2020.

\bibitem{liu2018backscatter}
W.~Liu, Y.-C. Liang, Y.~Li, and B.~Vucetic, ``Backscatter multiplicative
  multiple-access systems: Fundamental limits and practical design,''
  \emph{IEEE Trans. Wireless Commun.}, vol.~17, no.~9, pp. 5713--5728, 2018.

\bibitem{tse2005fundamentals}
D.~Tse and P.~Viswanath, \emph{Fundamentals of Wireless Communication}.\hskip
  1em plus 0.5em minus 0.4em\relax Cambridge university press, 2005.

\bibitem{cover1999elements}
T.~M. Cover, \emph{Elements of Information Theory}.\hskip 1em plus 0.5em minus
  0.4em\relax John Wiley \& Sons, 1999.

\bibitem{pillai2011capacity}
S.~R.~B. Pillai, ``On the capacity of multiplicative multiple access channels
  with {AWGN},'' in \emph{IEEE Inf. Theory Workshop}.\hskip 1em plus 0.5em
  minus 0.4em\relax IEEE, 2011, pp. 452--456.

\bibitem{smith1971information}
J.~G. Smith, ``The information capacity of amplitude-and variance-constrained
  scalar {Gaussian} channels,'' \emph{Inf. Control}, vol.~18, no.~3, pp.
  203--219, 1971.

\bibitem{shamai1995capacity}
S.~Shamai and I.~Bar-David, ``The capacity of average and peak-power-limited
  quadrature {Gaussian} channels,'' \emph{IEEE Trans. Inf. Theory}, vol.~41,
  no.~4, pp. 1060--1071, 1995.

\bibitem{dytso2019capacity}
A.~Dytso, M.~Al, H.~V. Poor, and S.~S. Shitz, ``On the capacity of the peak
  power constrained vector {Gaussian} channel: An estimation theoretic
  perspective,'' \emph{IEEE Trans. Inf. Theory}, vol.~65, no.~6, pp.
  3907--3921, 2019.

\bibitem{mamandipoor2014capacity}
B.~Mamandipoor, K.~Moshksar, and A.~K. Khandani, ``Capacity-achieving
  distributions in {Gaussian} multiple access channel with peak power
  constraints,'' \emph{IEEE Trans. Inf. Theory}, vol.~60, no.~10, pp.
  6080--6092, 2014.

\bibitem{ozel2012capacity}
O.~Ozel and S.~Ulukus, ``On the capacity region of the {Gaussian MAC} with
  batteryless energy harvesting transmitters,'' in \emph{Proc. of IEEE Glob.
  Commun. Conf. (GLOBECOM)}.\hskip 1em plus 0.5em minus 0.4em\relax IEEE, 2012,
  pp. 2385--2390.

\bibitem{liu2019intelligent}
F.~Liu, O.~Tsilipakos, A.~Pitilakis, A.~C. Tasolamprou, M.~S. Mirmoosa, N.~V.
  Kantartzis, D.-H. Kwon, J.~Georgiou, K.~Kossifos, M.~A. Antoniades
  \emph{et~al.}, ``Intelligent metasurfaces with continuously tunable local
  surface impedance for multiple reconfigurable functions,'' \emph{Phys. Rev.
  Appl.}, vol.~11, no.~4, p. 044024, 2019.

\bibitem{tang2020mimo}
W.~Tang, J.~Y. Dai, M.~Z. Chen, K.-K. Wong, X.~Li, X.~Zhao, S.~Jin, Q.~Cheng,
  and T.~J. Cui, ``{MIMO} transmission through reconfigurable intelligent
  surface: System design, analysis, and implementation,'' \emph{IEEE J. Sel.
  Areas Commun.}, vol.~38, no.~11, pp. 2683--2699, 2020.

\bibitem{zhao2021exploiting}
M.-M. Zhao, Q.~Wu, M.-J. Zhao, and R.~Zhang, ``Exploiting amplitude control in
  intelligent reflecting surface aided wireless communication with imperfect
  {CSI},'' \emph{IEEE Trans. Commun.}, vol.~69, no.~6, pp. 4216--4231, 2021.

\bibitem{wyner1966bounds}
A.~D. Wyner, ``Bounds on communication with polyphase coding,'' \emph{Bell
  Syst. Tech. J.}, vol.~45, no.~4, pp. 523--559, 1966.

\bibitem{el2011network}
A.~El~Gamal and Y.-H. Kim, \emph{Network Information Theory}.\hskip 1em plus
  0.5em minus 0.4em\relax Cambridge University Press, 2011.

\bibitem{goldsmith2003capacity}
A.~Goldsmith, S.~A. Jafar, N.~Jindal, and S.~Vishwanath, ``Capacity limits of
  {MIMO} channels,'' \emph{IEEE J. Sel. Areas Commun.}, vol.~21, no.~5, pp.
  684--702, 2003.

\bibitem{nocedal2014interior}
J.~Nocedal, F.~{\"O}ztoprak, and R.~A. Waltz, ``An interior point method for
  nonlinear programming with infeasibility detection capabilities,'' \emph{Opt.
  Methods SW.}, vol.~29, no.~4, pp. 837--854, 2014.

\bibitem{thangaraj2017capacity}
A.~Thangaraj, G.~Kramer, and G.~B{\"o}cherer, ``Capacity bounds for
  discrete-time, amplitude-constrained, additive white {Gaussian} noise
  channels,'' \emph{IEEE Trans. Inf. Theory}, vol.~63, no.~7, pp. 4172--4182,
  2017.

\bibitem{MathWorks}
\BIBentryALTinterwordspacing
``Constrained nonlinear optimization algorithms fmincon interior point
  algorithm,'' 2021. [Online]. Available: \url{Available: https://mathworks.
  com/help/optim/ug/constrained-nonlinear-optimization-algorithms.html}
\BIBentrySTDinterwordspacing

\bibitem{abou2001capacity}
I.~C. Abou-Faycal, M.~D. Trott, and S.~Shamai, ``The capacity of discrete-time
  memoryless rayleigh-fading channels,'' \emph{IEEE Trans. Inf. Theory},
  vol.~47, no.~4, pp. 1290--1301, 2001.

\bibitem{luenberger1997optimization}
D.~G. Luenberger, \emph{Optimization by Vector Space Methods}.\hskip 1em plus
  0.5em minus 0.4em\relax John Wiley \& Sons, 1997.

\bibitem{gradshteyn2014table}
I.~S. Gradshteyn and I.~M. Ryzhik, \emph{Table of Integrals, Series, and
  Products}.\hskip 1em plus 0.5em minus 0.4em\relax Academic press, 2014.

\end{thebibliography}
\end{document}